%% file: ms_astroph.tex
\documentclass[iop,revtex4]{emulateapj}
\slugcomment{{\sc Accepted to ApJ:} June 22, 2016} 

\pdfoutput=1

\usepackage{natbib}
\shorttitle{Molecular Gas in \iizw}
\shortauthors{Kepley et al.}

\newcommand{\ms}{\ensuremath{\rm{m \, s}^{-1}}}
\newcommand{\kms}{\ensuremath{\rm{km \, s}^{-1}}}
\newcommand{\Msun}{\ensuremath{\rm{M}_\odot}}

\newcommand{\uJybeam}{\ensuremath{\rm{\mu Jy \ beam^{-1}} \ }}
\newcommand{\mJybeam}{\ensuremath{\rm{mJy \ beam^{-1}} \ }}
\newcommand{\Jybeam}{\ensuremath{\rm{Jy \ beam^{-1}} \ }}
\newcommand{\co}[3]{\ensuremath{\rm{^{#1} CO ({#2}-{#3})}}}
\newcommand{\Kkms}{\ensuremath{\rm{K \, km \, s^{-1}}}}

\newcommand{\hii}{{\rm H\,}{{\sc ii}}}
\newcommand{\aco}{\ensuremath{\alpha_{co}}}
\newcommand{\Xco}{\ensuremath{{\rm X_{CO}}}}

\newcommand{\LCO}{\ensuremath{{\rm L_{CO}}}}
\newcommand{\LCOthree}{\ensuremath{{\rm L_{CO(3-2)}}}}
\newcommand{\LCOone}{\ensuremath{{\rm L_{CO(1-0)}}}}

\newcommand{\iizw}{II~Zw~40} \newcommand{\robzero}{robust~=~0}
\newcommand{\cothree}{\co{12}{3}{2}}
\newcommand{\cotwo}{\co{12}{2}{1}} \newcommand{\coone}{\co{12}{1}{0}}
\newcommand{\rthreetwo}{\ensuremath{{\rm r_{32}}}}
\newcommand{\rthreeone}{\ensuremath{{\rm r_{31}}}}
\newcommand{\rtwoone}{\ensuremath{{\rm r_{21}}}} 
\newcommand{\hi}{{\rm H\,}{{\sc I}}} 
\newcommand{\Kkmspc}{\ensuremath{\rm K \, \kms \,   pc^{2}}}
 \newcommand{\coh}{CO-to-H$_2$}
\newcommand{\MsunKkmspc}{\ensuremath{\rm M_\odot (\Kkmspc)^{-1}}}
\newcommand{\Msunpc}{\ensuremath{\rm M_\odot pc^{-2}}}
\newcommand{\gdr}{\ensuremath{\delta_{GDR}}}

\begin{document}



\title{The Molecular Clouds Fueling a 1/5 Solar Metallicity Starburst}

\author{Amanda A. Kepley} \affil{National Radio Astronomy Observatory,
  520 Edgemont Road, Charlottesville, VA 22903-2475}
\email{akepley@nrao.edu}

\author{Adam K. Leroy} \affil{The Ohio State University, McPherson
  Laboratory, 140 West 18th Avenue, Columbus, OH, 43210-1173} \email{leroy.42@osu.edu}
  
\author{Kelsey E. Johnson}
\affil{Department of Astronomy, University of Virginia, P.O. Box
  400325, Charlottesville, VA 22904-4325, USA}
\email{kej7a@virginia.edu}

\author{Karin Sandstrom}
\affil{University of California, San Diego, Center for Astrophysics \&
  Space Sciences, 9500 Gilman Drive, La Jolla, CA 92093-0424}
\email{kmsandstrom@ucsd.edu}

\author{C.-H. Rosie Chen}
\affil{Max-Planck-Institut für Radioastronomie, Auf dem Hügel 69, 53121 Bonn, Germany}
\email{rchen@mpifr-bonn.mpg.de}

\begin{abstract}

  Using the Atacama Large Millimeter/submillimeter Array, we have made
  the first high spatial and spectral resolution observations of the
  molecular gas and dust in the prototypical blue compact dwarf galaxy
  \iizw. The \cotwo\ and \cothree\ emission is clumpy and distributed
  throughout the central star-forming region. Only one of eight
  molecular clouds has associated star formation. The continuum
  spectral energy distribution is dominated by free-free and
  synchrotron; at 870\micron, only 50\% of the emission is from dust.
  We derive a \coh\ conversion factor using several methods, including
  a new method that uses simple photodissocation models and resolved
  CO line intensity measurements to derive a relationship that
  uniquely predicts \aco\ for a given metallicity. We find that the
  \coh\ conversion factor is 4 to 35 times that of the Milky Way (18.1
  to 150.5 ~\MsunKkmspc). The star formation efficiency of the
  molecular gas is at least 10 times higher than that found in normal
  spiral galaxies, which is likely due to the burst-dominated star
  formation history of \iizw\ rather than an intrinsically higher
  efficiency.  The molecular clouds within \iizw\ resemble those in
  other strongly interacting systems like the Antennae: overall they
  have high size-linewidth coefficients and molecular gas surface
  densities. These properties appear to be due to the high molecular
  gas surface densities produced in this merging system rather than to
  increased external pressure. Overall, these results paint a picture
  of \iizw\ as a complex, rapidly evolving system whose molecular gas
  properties are dominated by the large-scale gas shocks from its
  on-going merger.

\end{abstract}


\section{Introduction} \label{sec:introduction}

The high star formation rate surface densities and low metallicities
found in blue compact dwarf galaxies represent one of the most extreme
environments for star formation in the local universe, one more akin
to that found in high redshift galaxies than in local spirals
\citep{2009MNRAS.399.1191C,2011ApJ...728..161I}.  These global
properties result in increased disruption of the interstellar medium
by newly formed young massive stars
\citep{2006ApJ...653..361K,2010ApJ...709..191M}, higher and harder
radiation fields \citep{2006A&A...446..877M}, and reduced dust content
\citep{2013A&A...557A..95R}, all of which may significantly change how
the molecular gas within these galaxies transforms into stars.  To
date, however, the molecular gas fueling the starbursts within blue
compact dwarfs remains poorly understood due to the intrinsically
faint emission from the most common molecular gas tracers (CO and dust
continuum). By quantifying the properties of molecular gas in blue
compact dwarfs, we can determine how the physical conditions in these
galaxies influence their molecular gas, and thus the formation of
young massive stars, as well as gain insight into star formation in
high redshift galaxies, where detailed observations are difficult.

Previous low-resolution studies of molecular gas in low-metallicity
galaxies, including blue compact dwarfs, have shown that these
galaxies have extremely high star formation rates compared to their CO
luminosity -- a key tracer of the bulk molecular gas -- and that this
ratio increases as the metallicity of a galaxy decreases
\citep{1998AJ....116.2746T,2012AJ....143..138S}.  Taken by itself,
this trend suggests that low-metallicity galaxies either have
increased molecular star formation efficiency (i.e., less molecular
gas is necessary to form a given amount of stars) or that CO emission
is not as effective as tracer of molecular gas because of the
decreased dust shielding and reduced abundance of molecules in low
metallicity environments (i.e., less CO emission for a given amount of
molecular gas). Distinguishing between these two scenarios, however,
requires higher spatial resolution to directly measure the \coh\
conversion factor (and thus determine the total molecular gas mass)
and to link the young massive stars within the galaxy to the giant
molecular clouds from which they presumably form (although see
\citealp{2012ApJ...759....9K} and \citealp{2012MNRAS.421....9G} for
arguments that neutral hydrogen may play an increasing role in
star-forming clouds at low metallicity).

Resolved giant molecular cloud observations in low-metallicity
galaxies have historically been very difficult because of the faint
nature of the CO emission in these systems. One of the few resolved
studies of giant molecular clouds in a sample of dwarf galaxies showed
that the giant molecular clouds in these galaxies have similar sizes,
linewidths, and \coh\ conversion factors to more massive spiral
galaxies like the Milky Way, M33, and M31, which supports the idea of
higher star formation efficiencies \citep{2008ApJ...686..948B}. In
contrast, estimates of the \coh\ conversion factor from resolved dust
observations find systematically higher values for low-metallicity
galaxies, suggesting that dwarf galaxies have lower CO luminosities for
a given amount of molecular gas \citep{2011ApJ...737...12L}. These two
apparently contradictory sets of observations can be reconciled if the
reduced dust shielding for CO in low metallicity environments pushes
the CO emission to the densest portion of the molecular cloud, while the
H$_2$ remains distributed throughout the molecular cloud because it
can self-shield. Therefore, the CO observations only trace the central
regions of the molecular clouds, while the infrared observations trace
the dust, which is well-mixed with the surrounding envelope of molecular
gas
\citep{2008ApJ...686..948B,2011ApJ...737...12L,2013ApJ...777....5S}.

The sensitivity of the previous generation of millimeter
interferometers limited the sample in the most comprehensive study to
date \citep{2008ApJ...686..948B} to nearby galaxies ($\lesssim 4$ Mpc)
with relatively high metallicities; only one galaxy has a metallicity
less than 12+log(O/H)=8.2. Therefore, it is not surprising that these
authors see little variation in the sizes, linewidths, and \coh\
conversion factors of low metallicity galaxies. Intriguingly, they do
see hints that the lowest metallicity galaxy included in their sample
(the Small Magellanic Cloud) may deviate from the fiducial trends seen
in normal galaxies, although the deviations are relatively weak. This
result suggests that expanding resolved molecular gas studies to lower
metallicities and more extreme systems than possible with the previous
generation of millimeter interferometers may uncover more variations
in molecular cloud properties. Fortunately, today we have access to
the Atacama Large Millimeter/submillimeter Array (ALMA), whose
excellent sensitivity and resolution allow us to do just this.

The blue compact dwarf galaxy \iizw\ represents a key test case for
understanding how the properties of molecular clouds vary with
metallicity and star formation rate surface density: this galaxy
bridges the gap between ultra-low metallicity ($\sim 1/50 Z_\odot$)
starburst galaxies like SBS 0335-052 and I Zw 18 and starbursting
galaxies with normal solar metallicities. Although \iizw\ has only a
moderate metallicity (12+log(O/H)=8.09;
\citealp{2000ApJ...531..776G}), roughly comparable to the SMC, its
central star-forming region has an extraordinarily high star formation
rate surface density of 520~$\Msun \, {\rm yr^{-1} \, kpc^{-2}}$
\citep{2014AJ....147...43K}, comparable to that found in more massive
starburst galaxies.  Crucially for our purposes, this galaxy is also
only 10~Mpc away \citep{1988cng..book.....T}, which is two to five
times closer than other comparable galaxies like I~Zw~18 and
SBS0335-052. At this distance, we can resolve the giant molecular
cloud fueling the starburst within \iizw\ using only moderate angular
resolution -- 0.5\arcsec\ corresponds to 24~pc linear resolution --
allowing us to quantify the resolved properties of its molecular
gas. By comparing these properties to those in other starburst
galaxies of varying metallicity, we can begin to disentangle the
relative effects of metallicity and high star formation rate surface
density on the star formation efficiencies and CO luminosities in blue
compact dwarf galaxies.

In this paper, we present new, high spatial and spectral resolution
ALMA observations of the molecular gas and dust content of \iizw. Our
goal is to understand how the interplay of intense star formation and
low metallicity within this galaxy shapes its molecular gas and dust,
and whether dust and molecular gas in this galaxy differs in key ways
from that in other metal-rich star-forming galaxies. To do this, we
measure the properties of the dust and giant molecular clouds within
\iizw. We use these measurements to derive the \coh\ conversion factor
(\aco) -- which underlies most of what we know about star formation
beyond the Local Group -- in \iizw\ and compare it to other
observational and theoretical estimates for this factor. Then we
compare the molecular cloud properties in \iizw\ to the cloud
properties in other systems to see if there are systematic differences
in cloud properties with metallicity and/or star formation rate
surface density. Finally, we use these observations to place the
properties of the molecular gas and star formation within \iizw\ in
the larger context of star formation within galaxies.




\section{Data} \label{sec:data}

This paper presents ALMA Cycle 1 (proposal 2012.1.00984.S; PI:
A. Kepley) observations of the CO and continuum emission from the
central starburst region of the prototypical blue compact dwarf galaxy
\iizw\ at 3, 1\,mm, and 870\micron.  We describe the data
calibration and imaging below. Tables~\ref{tab:obs_summary} and
\ref{tab:spw} summarize the observations and spectral window
configurations.

\input{observation_summary} \input{correlator_setup}

The 870\micron\ and 1\,mm data were calibrated in CASA
\citep{2007ASPC..376..127M} using scripts provided by the North
American ALMA Science Center (NAASC). First, the water vapor
radiometer corrections and system temperatures were calculated and
applied to the data, and the flux density of the flux calibrator was
set. Then the bandpass solutions were calculated using the bandpass
calibrator. Next the amplitudes and phases of the calibration sources
were derived. The calibration sources had per-integration phases and
per-scan amplitudes derived and applied and were flux-scaled.  The
selected 870\micron\ phase calibrator was faint ($\sim$~0.06~Jy), so
we combined all four spectral windows and both polarizations to
increase the signal-to-noise of the solutions for that band.  Finally,
we applied the per-scan phases, per-scan amplitudes, and fluxscale for
the phase calibrator to the science target. The 3\,mm data were
calibrated using the CASA pipeline version Cycle2-R1-B, which follows
a similar calibration procedure.

All of the observations used quasars instead of planets to calibrate
the flux density scales, which introduces additional uncertainty into
the flux density scale.  For the 870\micron\ observations, the flux of
the ``flux'' calibrator (J0510+180) decreased by 30\% between the two
flux measurements bracketing our observations: its measured flux
density was $1.090 \pm 0.06$~Jy on 2013 September 29, but had dropped
to $0.710 \pm 0.04$~Jy by 2013 October 20 according to the ALMA Quasar
Catalogue \citep{2014Msngr.155...19F}. Extrapolating between these two
values gives a flux density for J0510+180 of 0.75~Jy. We used our
bandpass calibrator (J0421-120), which had a more stable flux density
value, to check the accuracy of our flux density scale. Adopting the
extrapolated flux density for J0510+180 yields a flux density within
5\% of its measured flux. For the 1\,mm observations, the flux of
J0510+180 at 1\,mm was measured 15 days after the observations to be
1.5~Jy. We used this measurement to set our flux density
scale. Unfortunately, the bandpass and phase calibrators did not have
recently measured flux densities, so we could not use them to
cross-check our flux density scale.  At 3\,mm, the flux calibrator
J0510+180 was observed within two days of our observations and had a
flux of 1.56 Jy. We conservatively assume that the flux density of our
observations is accurate to within 10\% for the 870\micron\ data and
within 20\% for the 3\,mm and 1\,mm data.

The data were not self-calibrated because the line and continuum
emission from the source was too faint. Self-calibration typically
requires a signal-to-noise of approximately 10 on an individual
baseline for times on the order of a few integrations. For both data sets,
the signal-to-noise on timescales of a few integrations was less than
4 for the line and continuum data.

We imaged the data in CASA. Since we are interested in measuring and
comparing the CO cloud properties, we imaged the line data sets with
\robzero\ to maximize angular resolution while retaining some
sensitivity to larger-scale emission. The line images were restricted
to the overlapping region of uv-space between the two data sets and
smoothed to the same beam size to ensure that we are comparing
emission on the same size scales for both lines.  We also created a
naturally weighted \cotwo\ image without any uv restrictions, which
will be the most sensitive to emission on large angular scales, to
compare to previous single-dish observations by
\citet{2004AaA...414..141A}. In addition to $^{12}$CO, there were
several additional isotopologues of CO in this data set (see
Table~\ref{tab:spw}). We note that the spectral window intended for
\co{13}{2}{1} missed the line and thus we did not produce a
\co{13}{2}{1} image. 

We also created a set of continuum images from both the ALMA data
presented in this paper and the VLA data from
\citet{2014AJ....147...43K}. We optimized for flux recovery and
matching uv distributions over resolution because the three different
continuum emission components -- synchrotron, free-free, and dust --
will have different extents. We imposed a inner uv plane limit of
16$k\lambda$ to match the largest angular scale and tapered the
naturally weighted data to a $1\arcsec$ resolution and restored the
data with a common 1\arcsec\ beam. This procedure provided the most
accurate match to the uv coverage and resolution between the
relatively heterogenous uv sampling at the different wavelengths,
although it is lower resolution than our line data cubes. For the VLA
data, the fluxes in these images are within 10\% of the fluxes
measured in \citet{2014AJ....147...43K}: -7\% for the 4.86GHz image,
-6\% for the 8.45 GHz image, and 1.5\% for the 22.46GHz image. Based
on the single-dish continuum measurements of
\citet{1991A&A...246..323K}, we only recover $\sim$50\% of the
emission from this galaxy and thus are resolving out continuum
emission on larger angular scales than we probe here
(12.4\arcsec). See the analysis in \S4.1 of
\citet{2014AJ....147...43K} for a detailed discussion of this point.

Tables~\ref{tab:image_summary_line} and \ref{tab:image_summary_cont}
give a summary of our image properties. The largest angular scale of
these images is approximately 12.4\arcsec.

\input{image_summary_line.tex}

\input{image_summary_cont.tex}

\section{Results} \label{sec:results}

\subsection{Overview of the Data} \label{sec:overview-data}

\begin{figure*}
\centering
\includegraphics[width=\textwidth]{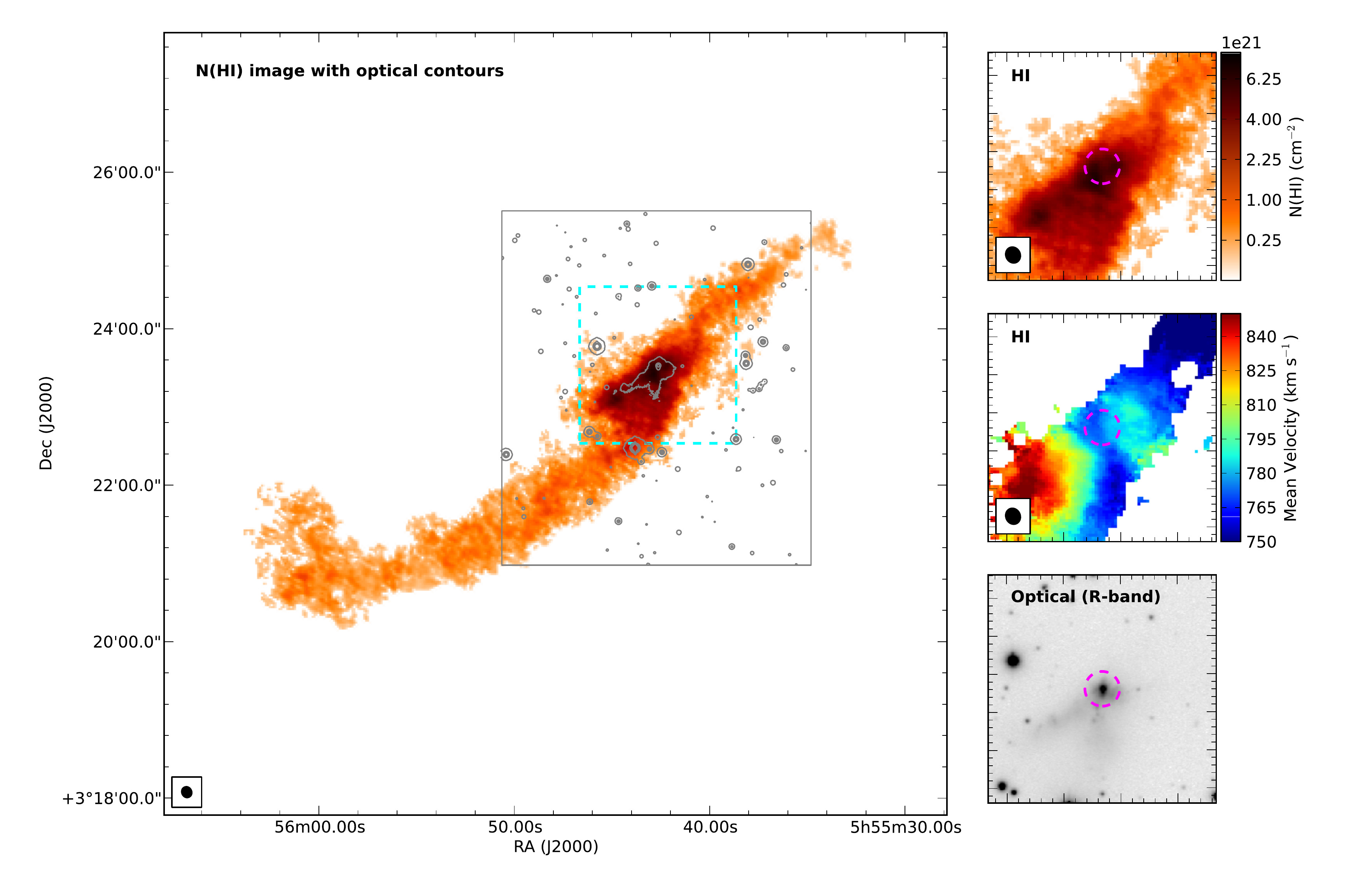}
\caption{The central starburst in \iizw\ was likely triggered by a
  major merger between two dwarf galaxies as evidenced by its extended
  neutral hydrogen tails and compact optical emission. {\em Left:}
  Neutral hydrogen column density image \citep{1998AJ....116.1186V}
  with contours showing the R-band optical emission with the gray box
  showing the region observed in the optical
  \citep{2003ApJS..147...29G}. The tidal tails are seen in both
  neutral hydrogen and in the R-band optical image, although the
  neutral hydrogen tidal tails extend several times further than those
  seen in the optical.  {\em Right:} A closer look at the neutral
  hydrogen and optical emission in the region outlined by the cyan box
  in the left panel. The right panels show (top to bottom) the neutral
  hydrogen column density, the neutral hydrogen mean velocity, and an
  optical R-band image. The field of view for the 870\micron\ ALMA
  observations presented in this paper is shown as a magenta
  circle. These observations probe a region of relatively high HI
  column density ($5 \times 10^{21} \, {\rm cm}^{-2}$) and a region
  where the HI velocity field undergoes a reversal. The optical
  nucleus is compact, as one would expect from an evolved merger
  remnant.  The beam for the neutral hydrogen observations is shown in
  the lower left-hand corner of the relevant panels.}
\label{fig:iizw40_overview}
\end{figure*}

The blue compact dwarf galaxy \iizw\ consists of a central starburst
region, visible as a compact ($\sim$8\arcsec\ = 390pc) knot in the
optical, embedded within an large neutral hydrogen envelope
characterized by extended tidal tails
(Figure~\ref{fig:iizw40_overview}). Given its morphology, the
starburst is the likely result of a major merger between two dwarf
galaxies \citep{1982MNRAS.198..535B,1998AJ....116.1186V}. The burst
hosts at least three young massive clusters brighter than 30 Doradus
\citep{2014AJ....147...43K}, but little is known about the molecular
gas that is presumably fueling this extraordinary episode of star
formation.

To quantify the properties of the molecular gas fueling the central
starburst within \iizw, we used ALMA to observe \coone, \cotwo, and
\cothree\ as well as the continuum at 3\,mm, 1\,mm, and 870\micron\ in
a single field centered on the central starburst region, indicated by
a magenta circle in the rightmost panels of
Figure~\ref{fig:iizw40_overview}. Using high-resolution radio
continuum observations, the star formation rate surface density in
this region was found to be 520~$\Msun \, {\rm yr^{-1} \, kpc^{-2}}$
\citep{2014AJ....147...43K}: five orders of magnitude higher that of
the solar neighborhood Milky Way value
(0.003~$\Msun \, {\rm yr^{-1} \, kpc^{-2}}$) and the same order of
magnitude as star formation rates in more massive starburst systems
like LIRGs and ULIRGs
\citep[100-1000~$\Msun \, {\rm yr^{-1} \,
  kpc^{-2}}$;][]{1998ApJ...498..541K}.
This region also has high \hi\ surface densities
($5\times10^{21} {\rm cm}^{-2}$) consistent with the threshold for
${\rm H_2}$ formation in low metallicity galaxies
\citep{2010ApJ...709..308M,2013ApJ...777L...4W} and is located near an
\hi\ velocity reversal (see the middle panel on the right-hand side of
Figure~\ref{fig:iizw40_overview}).

\begin{figure}
\centering
\includegraphics[width=\columnwidth]{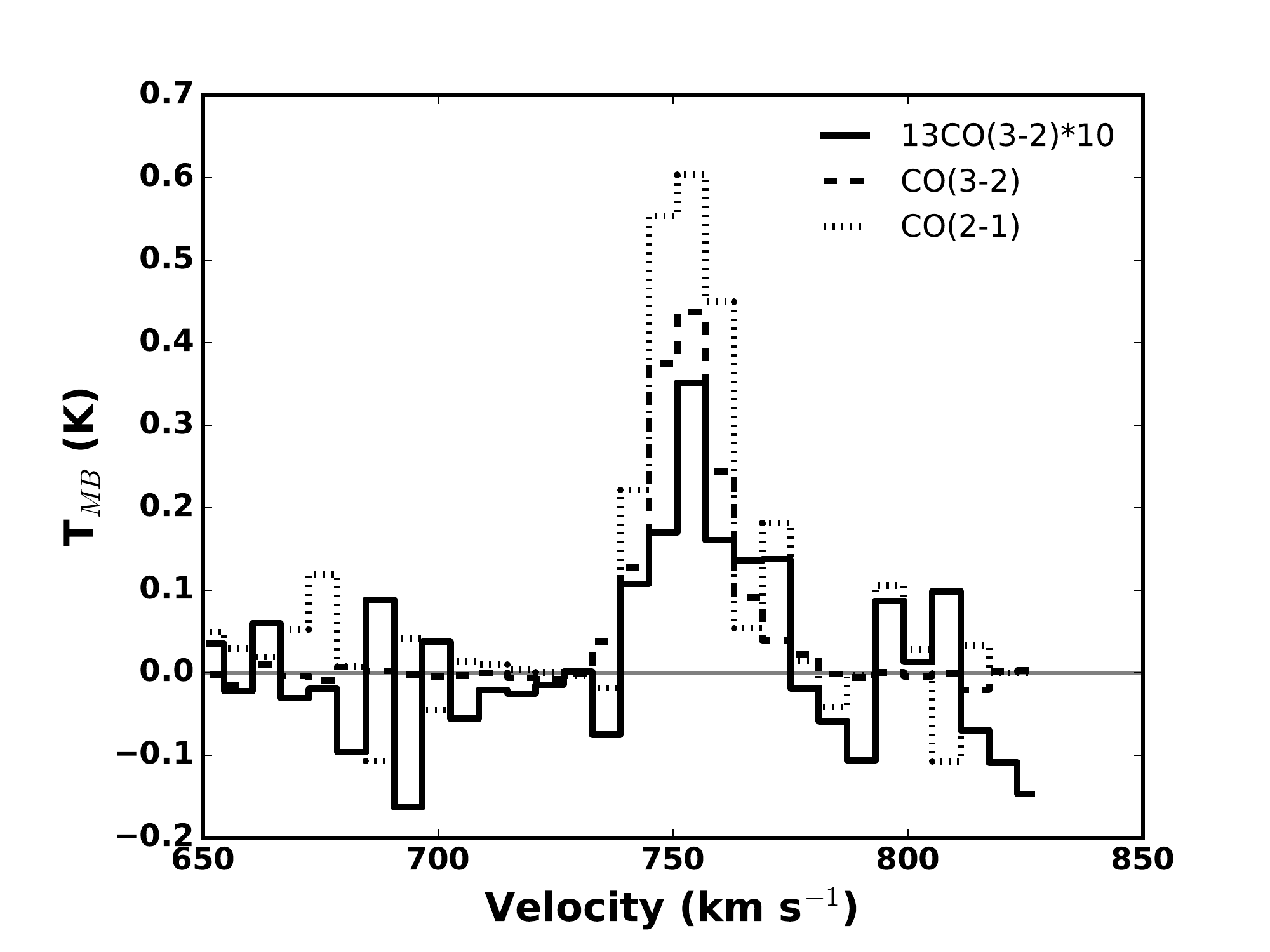}
\caption{Average \co{12}{3}{2}, \co{12}{2}{1}, and \co{13}{3}{2}
  spectra over the main CO ridge (extraction region corresponds to the
  region labeled BCD in Figure~\ref{fig:co_spectra}). We clearly
  detect \co{13}{3}{2} emission in the average spectrum, but do not
  detect it in the individual channel maps.} \label{fig:13co32}
\end{figure}

Our observations detected \cotwo\ and \cothree\ emission from the
central region as well as continuum emission at all three
wavelengths. We also have a global detection of \co{13}{3}{2} when we
average over the main ridge of CO emission
(Figure~\ref{fig:13co32}). The \cotwo\ emission generally has a higher
brightness temperature than the \cothree\ emission, but the \cothree\
observations have better brightness temperature sensitivity
($\sigma_{T_B} \propto \sigma_{S_\nu}/ \nu^2$), leading to higher
signal-to-noise for the \cothree\ data
(Figure~\ref{fig:co_spectra}). The \coone\ line was not
detected. However, given the higher noise in the \coone\ image (in K),
this non-detection is consistent with the ratio between the \cotwo\
and \coone\ emission seen in \iizw\ in other lower-resolution
observations \citep[0.5;][]{2004AaA...414..141A}.

\subsubsection{CO emission in \iizw} \label{sec:co-emission-iizw}

\begin{figure*}
\centering
\includegraphics[width=\textwidth]{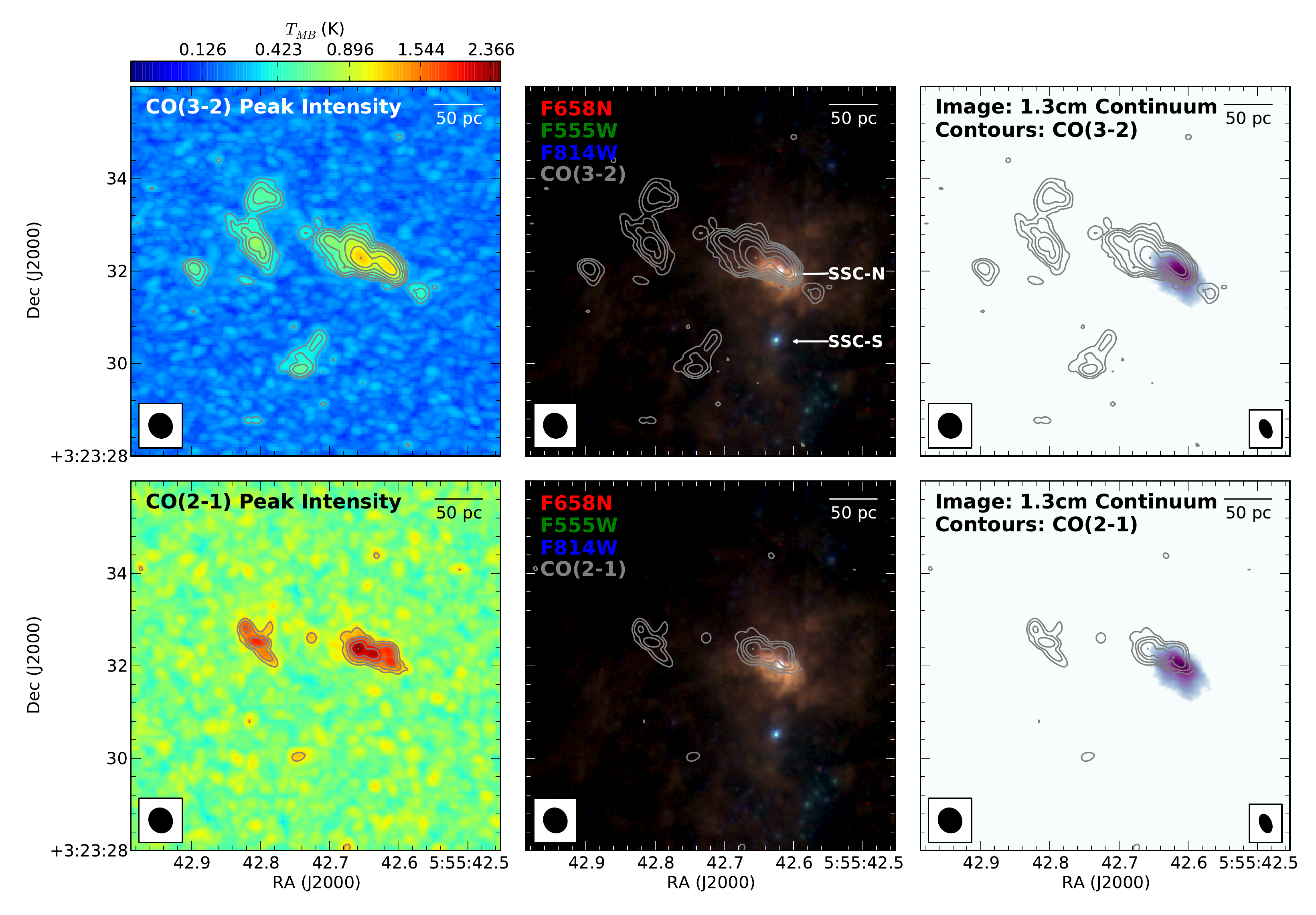}
\caption{The \cothree\ and \cotwo\ emission detected by ALMA is clumpy
  and distributed throughout the central starburst region. Only 30\%
  of the CO emission is coincident with star formation. {\em Left
    Column:} Peak intensity maps (in units of $T_{MB}$) for \cothree\
  (top) and \cotwo\ (bottom). Both images have the same resolution and
  are on the same color scale. The \cothree\ contours start at 0.31~K
  and increase by factors of 1.3 and the \cotwo\ contours start at
  1.16~K and increase by factors of $\sqrt{2}$.  These images show
  that the \cotwo\ emission has a higher brightness temperature than
  the \cothree\ emission, but the \cothree\ data has less noise,
  giving the \cothree\ data a higher signal to noise ratio.  {\em
    Middle Column:} The contours for \cothree\ (top) and \cotwo\
  (bottom) emission overlaid on a three-color {\em Hubble Space
    Telescope (HST)} image showing the optical emission in the central
  starburst region of \iizw\ \citep{2014AJ....147...43K}. This region
  hosts a number of young massive clusters and has a significant
  amount of ionized gas emission. The two super star clusters
  analyzed in \citet{2014AJ....147...43K} are labeled. {\em Right
    Column:} Image showing the 1.3\,cm continuum, which is dominated
  by free-free emission from young massive stars
  \citep{2014AJ....147...43K}, with \cothree\ (top) and \cotwo\
  (bottom) contours overlaid.  Only the western end of the main CO
  region has associated 1.3\,cm emission.  We note that the region
  shown has been choosen to highlight the structure of the molecular
  gas in the central starburst. One cloud (Cloud A in the terminology
  of \S~\ref{sec:meas-cloud-prop}) falls outside. This cloud can be
  seen in the lower right hand corner of
  Figure~\ref{fig:co_spectra}. }
\label{fig:co_plus_stars}
\end{figure*}

\begin{figure*}
\centering
\includegraphics[width=\textwidth]{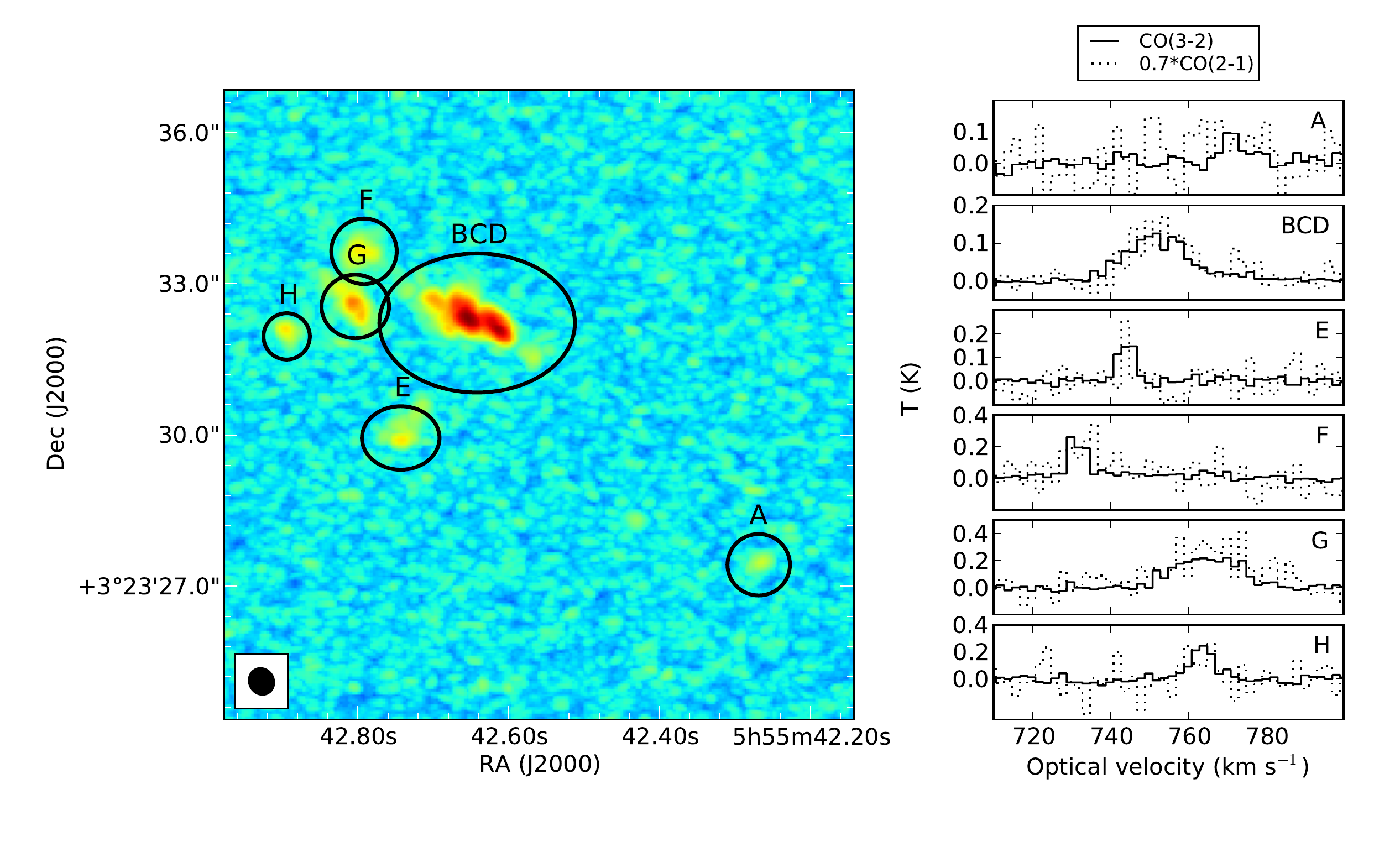}
\caption{\cothree\ and \cotwo\ spectra from different regions in
  \iizw.  On the left side of the image, the extraction regions are
  shown overlaid on a \cothree\ image. These regions roughly
  correspond to the clouds identified in \S~\ref{sec:meas-cloud-prop}
  and are labeled accordingly. The \cothree\ and \cotwo\ spectra from
  each region are shown on the right. In general, the \cotwo\ emission
  is stronger than the \cothree\ in \iizw, but the \cothree\
  observations have higher signal to noise. The \rthreetwo\ brightness
  temperature ratio is 0.7 for most regions, although cloud G has a
  lower \rthreetwo\ ratio of 0.5.}
\label{fig:co_spectra}
\end{figure*}

The observed \cotwo\ and \cothree\ emission is clumpy and distributed
throughout the central star-forming region of \iizw\
(Figure~\ref{fig:co_plus_stars}).  This CO distribution can be compared to
distribution of young massive clusters traced by the 1.3\,cm radio
continuum emission from \citet{2014AJ....147...43K}. These deep,
high-resolution observations detect clusters with masses as low as
$1.5\times10^4\,\Msun$ anywhere within the \cothree\ field of
view.\footnote{Note that Figure 2 in \citet{2014AJ....147...43K} does
  not show the entire field of view for these observations.} We find
that only the western end of the main CO ridge is coincident with
1.3\,cm continuum emission indicating recent star formation. The other
70\% of the CO emission has no coincident 1.3\,cm\ emission.

The morphology of the molecular gas emission in \iizw\ strongly
resembles that found in the nearby dwarf starburst galaxy NGC 1569
\citep{1999A&A...349..424T}, which is at a distance of 3.36 Mpc
\citep{2008ApJ...686L..79G}. In that galaxy, there are several
well-separated molecular gas clouds. The molecular gas is not
spatially associated with the two older ($>7$Myr) super star clusters
found in the center of NGC 1569 \citep{2000AJ....120.2383H}. However,
the tip of one cloud is associated with an \hii\ region. This
molecular gas distribution is in contrast to that found in
observations with similar resolution of more massive (and metal-rich)
starburst galaxies such as M82 \citep{2005ApJ...635.1062K} and the
Antennae \citep{2012ApJ...750..136W}. In these galaxies, the molecular
gas emission is much more extensive and complex, although in the case
of M82 this may be more a result of the high inclination of this
galaxy rather than the intrinsic molecular gas
distribution. Separation of the molecular gas into individual
molecular clouds for both the Antennae and M82 is much more difficult,
and individually identified clouds frequently show signs of having
multiple components along the line of sight
\citep{2005ApJ...635.1062K,2012ApJ...750..136W}. In both galaxies, the
star formation appears to be concentrated on the edges of the
molecular gas clouds \citep{2005ApJ...635.1062K,2012ApJ...750..136W}.

The absence of 1.3\,cm continuum emission from most of the CO clouds
is unlikely to be the effect of extinction. Once emitted the 1.3\,cm
emission is unaffected by dust. We note, however, that dust can still
absorb photons before they ionize the surrounding gas and produce
free-free emission. This effect is roughly on the order of 30\% for
the Large Magellanic Cloud \citep{2001AJ....122.1788I} and decreases
as a function of metallicity.  It is also unlikely to be star
formation below the 1.3\,cm detection limit. The least massive CO
cloud in our sample (Cloud E in Figure~\ref{fig:cprops_clumps}) has a
mass of $\sim2\times10^5\,\Msun$, assuming the minimum \aco\ derived
for \iizw\ (see \S~\ref{sec:coh-ratio-molecular} for details). For a
typical star formation efficiency of 3\% \citep{2010ApJ...723.1019H},
this implies a stellar mass of $\sim0.5\times10^4\,\Msun$ which is at
the 1.3\,cm detection limit. Higher \aco\ values for \iizw, which are
likely, would imply even larger cluster masses. All this evidence
suggests that the current star formation within \iizw\ is associated
only with the western edge of the main CO ridge and that the rest of
the molecular gas does not appear to be forming stars.

Given GMC-scale ($\sim 24$~pc) resolution of the \cothree\ and
\cotwo\ data, we do not expect all the molecular clouds within \iizw\
to have associated star formation. On large ($\sim$ kpc) scales, the
star formation rate surface density is strongly correlated with the
molecular gas surface density
\citep{2002ApJ...569..157W,2013AJ....146...19L}. This relationship
breaks down on smaller GMC-sized ($\sim 10-75$pc) scales because the
observations are averaging over fewer molecular clouds, accentuating
the evolutionary state of individual clouds
\citep{2010ApJ...722.1699S,2010ApJ...722L.127O,2010ApJ...721.1206C}.

Using the classification system of \citet{2009ApJS..184....1K}, we can
classify the molecular clouds within \iizw\ into three different
types.  Type I clouds are the youngest molecular clouds and show no
sign of current star formation. Type I clouds evolve into Type II
clouds, which have \hii\ regions generated by the ionizing radiation
from young massive clusters, but no visible optical clusters. Type III
clouds represent the final evolutionary stage in a molecular cloud's
life. They have both \hii\ regions and visible optical clusters and
are being destroyed by their embedded star formation. For the clouds
in the LMC, \citet{2009ApJS..184....1K} estimated lifetimes for three
types of clouds of 6Myr, 13Myr, and 7Myr, assuming a constant star
formation rate. This classification system is analogous to the scheme
developed by \citet{2014ApJ...795..156W} to classify molecular clouds
within the Antennae galaxies.

In \iizw, cloud C would be classified as a Type III cloud. The
remainder of the molecular clouds within \iizw\ are Type I
clouds. Using optical data, \citet{2014AJ....147...43K} determined
that the optical cluster associated with the western end of the
molecular gas ridge is less than 5~Myr old, which is consistent with
the lifetime of Type III clouds in the LMC estimated by
\citet{2009ApJS..184....1K}. The optical cluster to the south of the
main star forming region is 9.5~Myr old and shows no sign of
associated molecular gas, which is again consistent with the timescale
for Type III clouds in the LMC.

\iizw\ has three times more Type I clouds than one would predict based
on the relative numbers of clouds in the LMC (7 instead of
2). Although we are in the realm of small number statistics, this
difference is still significant. This preponderance of Type I clouds
has several possible causes. First, the lifetimes of the Type I clouds
in \iizw\ could be longer than similar clouds in the LMC. Second, the
Type I clouds all formed at the same time but have not yet had time to
form stars.  Third, these clouds are stochastic, unbound,
agglomerations of molecular gas that will disperse in the
future. Finally, any Type II and III clouds could have been rapidly
destroyed via feedback from the central starburst within \iizw. Clouds
in the process of destruction could also be fainter than Type I clouds
and thus may be below our detection limit. Given the evolutionary
history of \iizw, we suggest that the latter three possibilities are
more likely, although we do not have enough evidence to definitely say
which mechanism is at work.

For the entire galaxy, the CO line brightness temperature ratios are
consistent with an optically thick blackbody with an excitation
temperature of 10K, i.e., an \rthreetwo\ value of 0.7, \rthreeone\
value of 0.54, and \rtwoone\ value of 0.74. From our 870\micron\ and
1\,mm cubes, we find an \rthreetwo\ ratio of 0.7 with an uncertainty
of 30\% for the entire region; the large uncertainty in this value is
due mainly to the overall uncertainty in the flux density
calibrations.  Adjusting for the relative beam sizes, single-dish
observations by \citet{2004AaA...414..141A} and
\citet{2001AJ....121..740M} found \rtwoone\ values of 0.5 and
$0.58\pm 0.16$, respectively. Both of these values are within the
error of the \rtwoone\ value we derive from our data. However, we note
that the single dish observations will sample different mean molecular
gas conditions from the higher resolution observations presented here,
so exact correspondence is not expected. In general, the line ratios
are not consistent with optically thin gas in local thermodynamic
equilibrium (LTE). In that case, the line ratios would be greater than
one, except for extremely low excitation temperatures ($\sim$few
K). The \co{12}{3}{2} to \co{13}{3}{2} ratio averaged over the main CO
ridge is $15\pm 12$, which is consistent with optically thin
\co{13}{3}{2} and optically thick \co{12}{3}{2}.

For the above arguments, we have assumed that the gas is in
LTE. However, the CO in \iizw\ does not necessarily have to be in
LTE. If the gas does not have sufficiently high density, we could have
line ratios that mimic the ratios from optically thick emission (less
than unity) because the upper levels are not populated. Given that all
other evidence suggests we are observing dense star-forming emission
cores, we view this as an unlikely scenario.

The line ratios within \iizw\ are consistent with the observed line
ratios derived from single dish observations of a larger sample of
dwarf starburst galaxies (including \iizw) by
\citet{2001AJ....121..740M}. That work found a mean, error-weighted
value of \rthreeone\ of $0.60\pm0.06$, which is more similar to that
found in LIRGs \citep[0.47; 0.1-0.7][]{2010MNRAS.406.1364L} and in the
centers of galaxies \citep[0.2 to 0.7 and 0.61,
respectively][]{1999A&A...341..256M,2010ApJ...724.1336M} than in
nearby normal star-forming galaxies
\citep[$0.18\pm0.02$;][]{2012MNRAS.424.3050W}.

\begin{figure}
\centering
\includegraphics[width=\columnwidth]{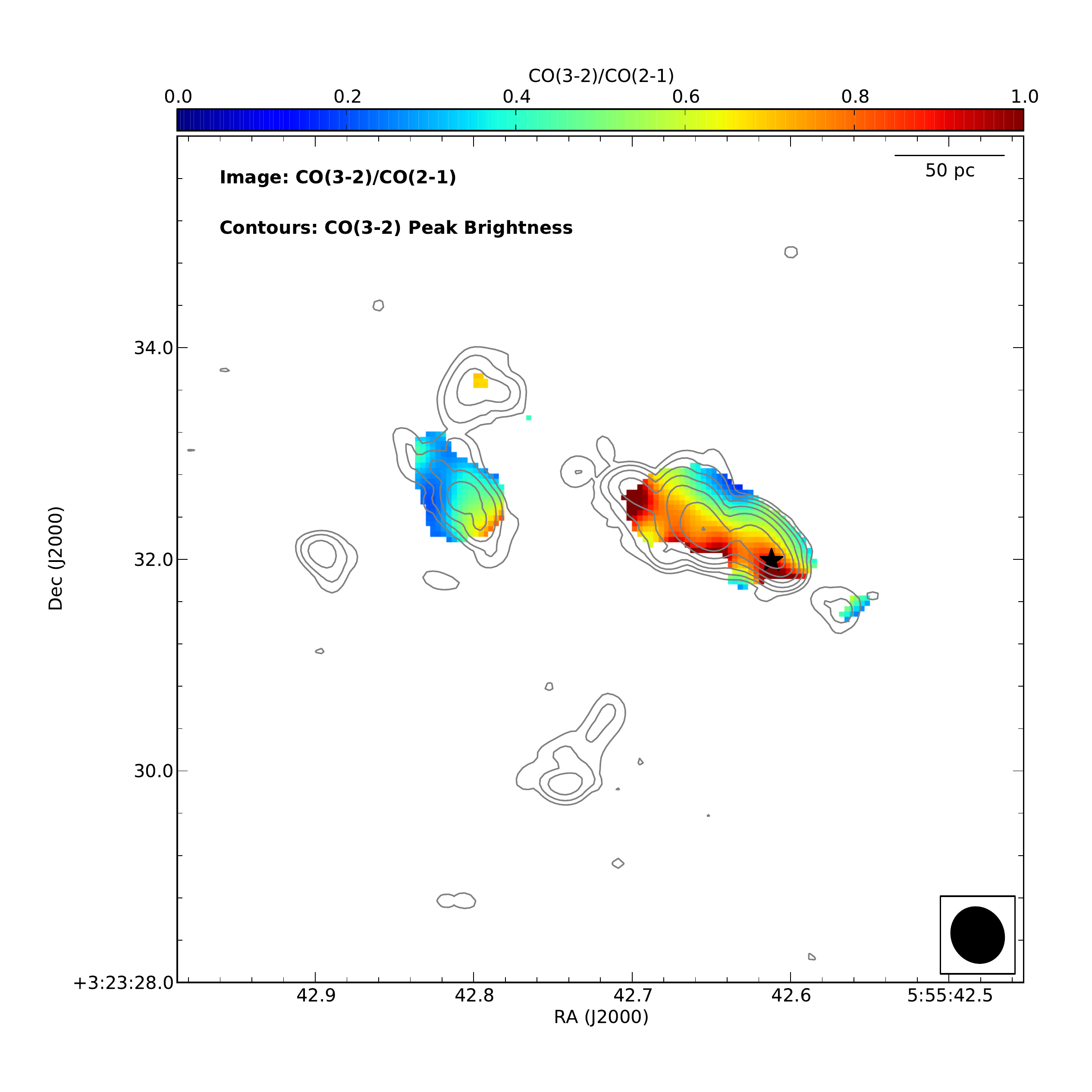}
\caption{The image shows the \rthreetwo\ brightness temperature ratio
  with contours showing \cothree\ peak brightness temperature from
  Figure~\ref{fig:co_plus_stars} overlaid. The location of the 1.3\,cm
  continuum peak is indicated by a star. The \rthreetwo\ brightness
  temperature ratio is higher closer to the actively star-forming
  region in the center of \iizw.}
\label{fig:iizw40_r32}
\end{figure}

Although the \rthreetwo\ ratio is relatively constant across the face
of the galaxy, there are two outliers: cloud G and the western edge of
the main CO ridge. The \rthreetwo\ line brightness temperature ratio
for cloud G (0.53) is consistent with a blackbody with a lower
excitation temperature of $\sim$ 6K, which implies an \rthreeone\
value of 0.34.  The western edge of the main CO ridge has elevated
\rthreetwo\ ratios roughly a factor of two, indicating higher CO
excitation, which is consistent with presence of star formation in
this region (Figure~\ref{fig:iizw40_r32}).

\begin{figure*}
\centering
\includegraphics[width=\textwidth]{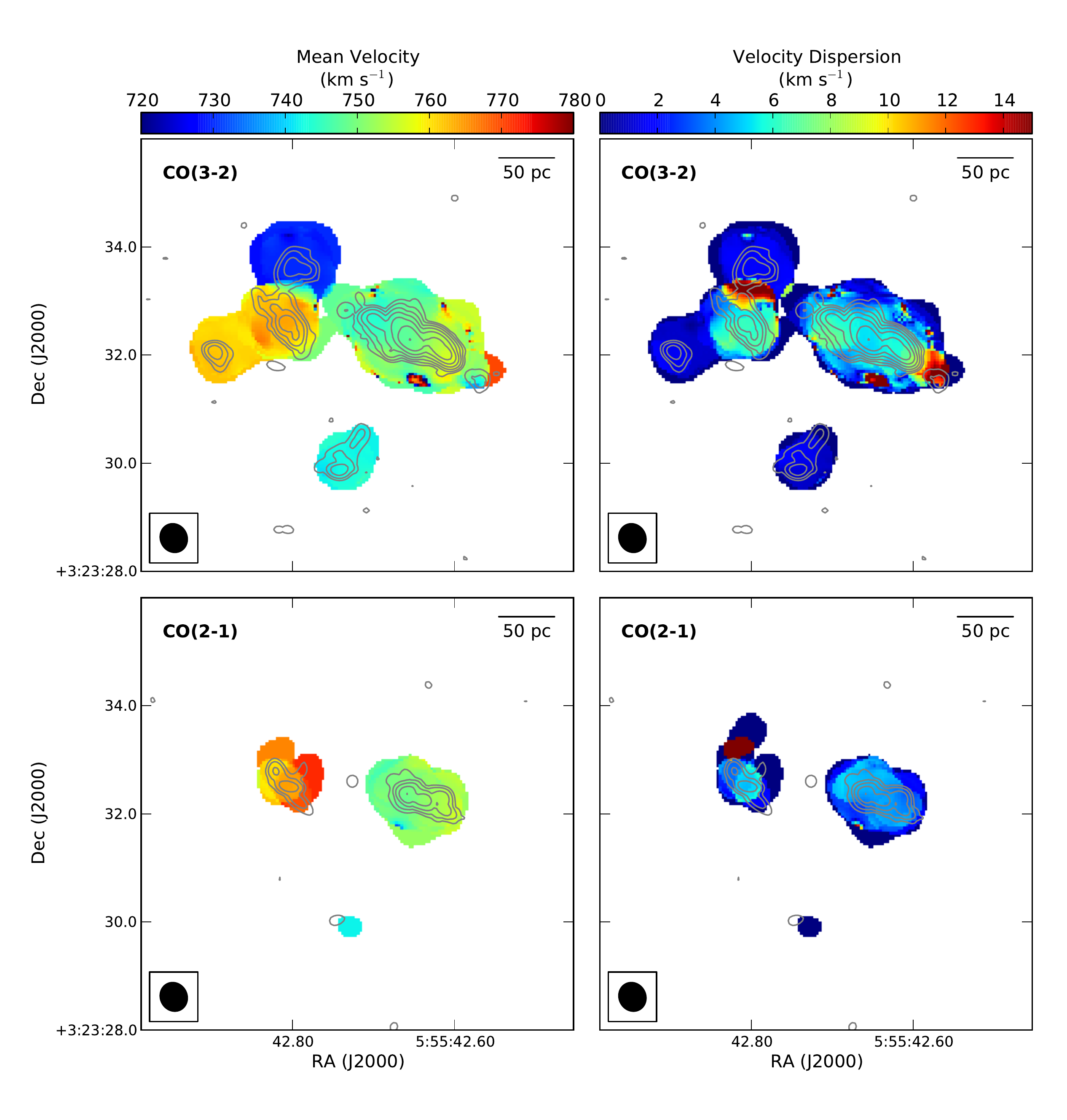}
\caption{ {\em Left column:} The mean velocity for the \cothree\ (top)
  and \cotwo\ (bottom) emission. The peak brightness temperature
  contours from Figure~\ref{fig:co_plus_stars} are overlaid. {\em
    Right column:} The right column shows the velocity dispersion for
  the \cothree\ (top) and \cotwo\ (bottom) emission.  The peak
  brightness temperature contours from Figure~\ref{fig:co_plus_stars}
  are overlaid. The velocity field for \iizw\ shows little evidence
  for large scale rotation, except for the main CO emitting
  region. The velocity dispersions are highest in the brightest
  emitting CO clumps. In particular, the region near 05:55:42.8,
  03:23:33, has an extremely high line width compared to the other
  regions ($\sim$6 \kms\ vs. $\sim$1-2 \kms). The $\sim 10-14\kms$
  velocity widths just to the north of this region are due to the
  overlap of two different clouds. See Figure~\ref{fig:cprops_clumps}
  and \S~\ref{sec:meas-cloud-prop} for a discussion. We note that the
  region shown has been chosen to highlight the structure of the
  molecular gas in the central starburst. One cloud (Cloud A in the
  terminology of \S~\ref{sec:meas-cloud-prop}) falls outside. This
  cloud can be seen in the lower right hand corner of
  Figure~\ref{fig:co_spectra}.}
\label{fig:co_moments}
\end{figure*}

The velocity field for \iizw\ does not show strong evidence for
rotation except for the main ridge of CO emission
(Figure~\ref{fig:co_moments}). The rest of the CO clouds appear to be
relatively well separated in velocity space. Most of the clouds have
velocity dispersions of $\sim$2 \kms, except for the two bright
emission peaks, which have larger velocity dispersions of $\sim$6
\kms. 

\begin{figure*}
\centering
\includegraphics[width=\textwidth]{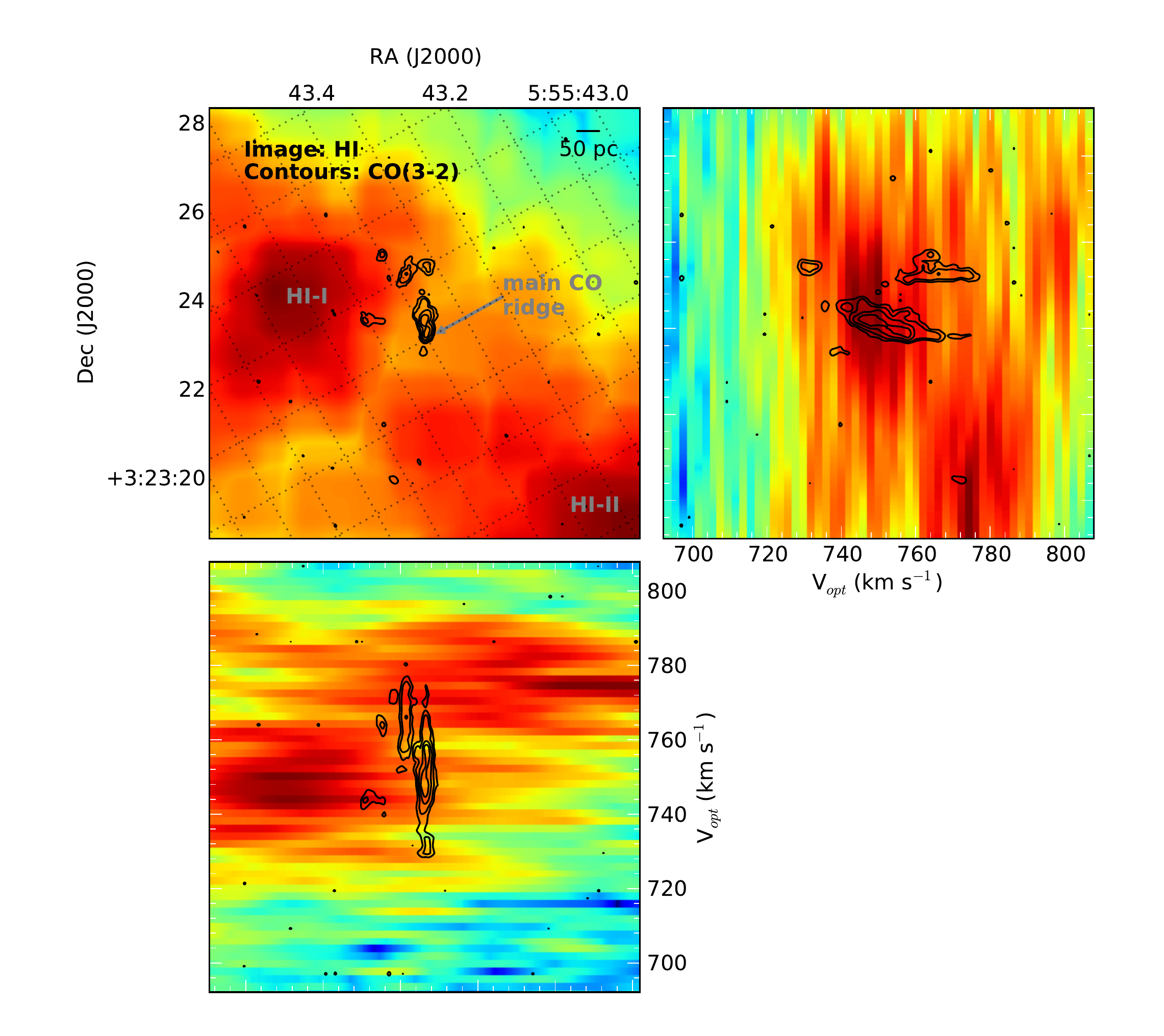}
\caption{A comparison of the \hi\ and \cothree\ data cubes. The cube
  has been rotated by 61 degrees so that the main ridge of CO emission
  runs up and down. The CO contours start at the three sigma noise
  level (0.087 \Jybeam) and increase by powers of two (0.174, 0.348,
  0.696, \ldots \Jybeam). The \cothree\ emission lies between two \hi\
  peak (referred to here at HI-I and HI-II) and the velocities of the
  individual regions are coincident with the velocities of both \hi\
  peaks. The spatial distribution of the emission combined with the
  overlapping velocities suggests that the starburst in \iizw\ may
  have been triggered by colliding gas clouds. }
\label{fig:co_hi_pv}
\end{figure*}

The \cothree\ velocity field overlaps, but is not completely
coincident with the \hi\ velocity field
(Figure~\ref{fig:co_hi_pv}). In the plane of the sky, the CO emission
lies between two \hi\ peaks, which we refer to as HI-I and HI-II. In
velocity space, the CO clouds overlap the velocities of both peaks,
although the CO main ridge appears to be more closely associated with
the neutral hydrogen peak HI-I and the high velocity disperion region
with HI-II. 

Although \iizw\ has not had its large-scale kinematics modeled in
detail, its \hi\ velocity field bears a strong resemblance to that of the
merging galaxy NGC~7525, which has been modeled in detail by
\citet{1995AJ....110..140H}. In particular, the \hi\ velocity reversal
within the field of view of the CO observations is consistent with gas
from the progenitor galaxies flowing inward (c.f., Figure 6 in
\citet{1995AJ....110..140H}, especially T=32 and T=40). It also agrees
with the scenario presented in \citet{1988MNRAS.231P..63B} that the
infalling gas from the progenitor systems may have led to the
formation of molecular gas in and thus triggered the current burst of
star formation in \iizw. High resolution N-body/SPH simulations of the
merging galaxy Antennae have shown that a merger can produce an
short-lived, off-center burst of star formation
\citep{2010ApJ...715L..88K}.

\begin{figure}
\centering
\includegraphics[width=\columnwidth]{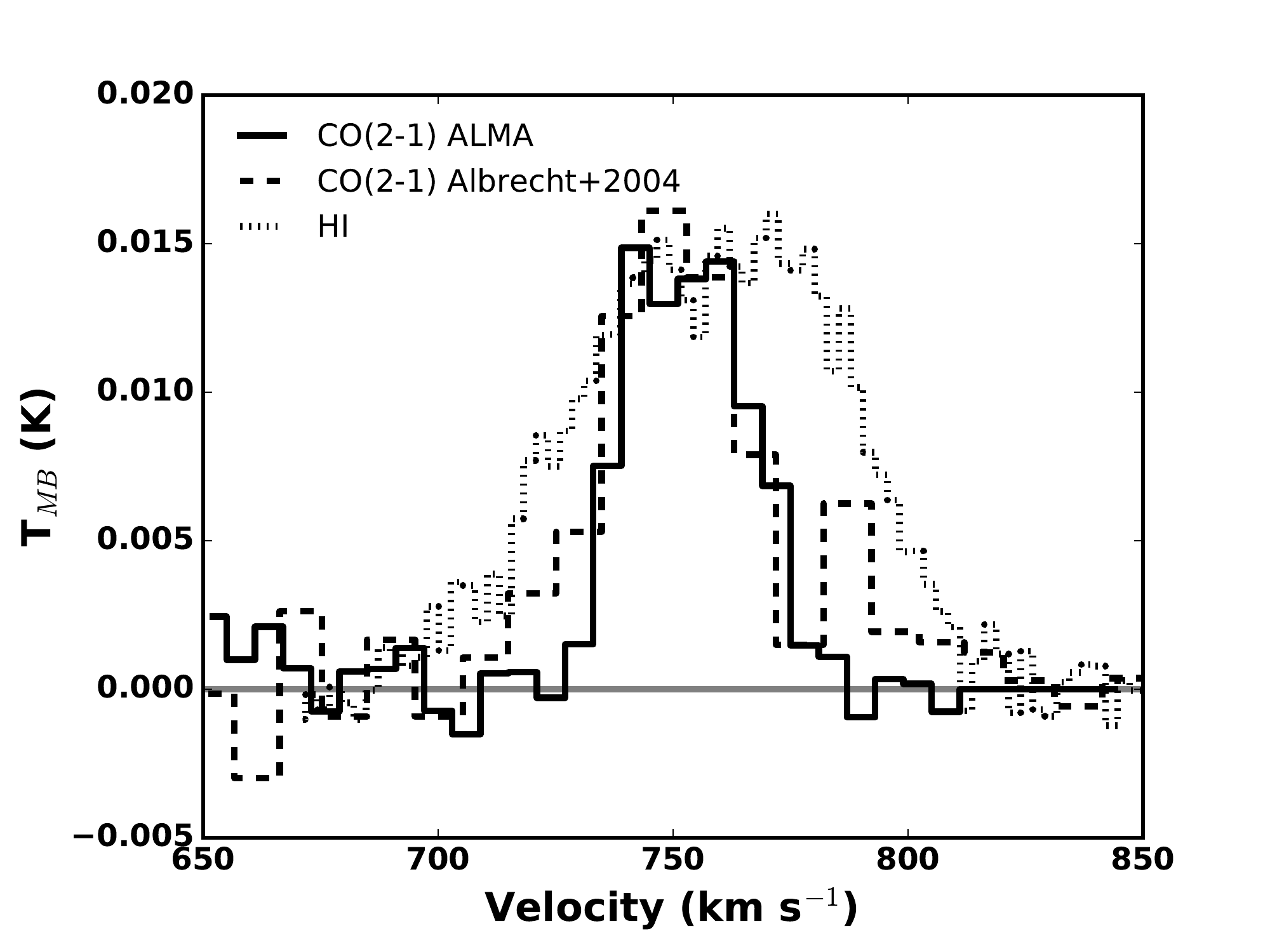}
\caption{A comparison of the ALMA \cotwo\ spectrum (solid black line)
  to a single dish \cotwo\ spectrum from the IRAM 30m (dashed line;
  \citealt{2004AaA...414..141A}) shows that 38\% of the \cotwo\
  emission is resolved out of the ALMA observations.  Although both
  profiles have similar central velocities, the ALMA \cotwo\ profile
  is missing emission in the line wings suggesting that there is
  diffuse CO emission at non-systemic velocities. Both \cotwo\
  profiles peak near the peak of the neutral hydrogen profile (dotted
  line, maximum height scaled to 80\% of the plot size), but have much
  narrower line widths.}
\label{fig:total_spectrum}
\end{figure}

Approximately 30\% of the total \cotwo\ emission in \iizw\ is resolved
out by our ALMA observations. The integrated \cotwo\ intensity for
\iizw\ from single dish observations with the IRAM 30m is 0.69~\Kkms\
\citep{2004AaA...414..141A}, while the integrated \cotwo\ intensity
for the ALMA \cotwo\ cube smoothed to the same resolution (11\arcsec)
is 27\% lower (0.5~\Kkms;
Figure~\ref{fig:total_spectrum}).\footnote{The size of the single dish
  beam is 11\arcsec, which is comparable to the largest angular scale
  observed by ALMA in this image.}  Although both the single dish and
ALMA \cotwo\ profiles have similar central velocities, the ALMA
\cotwo\ profile is missing emission in the line wings, suggesting that
there is diffuse CO emission at non-systemic velocities. This result
is consistent with CO observations in galaxies, in particular the
Local Group dwarf galaxy IC 10 \citep{2006ApJ...643..825L}.

\subsubsection{Continuum Emission in \iizw} \label{sec:cont-emiss-iizw}

Dust is key to our picture of how CO emission changes with
metallicity. As mentioned in the introduction, dust shields CO from
disassociation, while ${\rm H_2}$ is able to self-shield. The reduced
dust content found in low metallicity galaxies means less dust is
available to shield CO, while the molecular gas itself is unaffected
because H$_2$ self-shields. To explore the interplay of dust and CO
emission in \iizw, we compile a submillimeter to centimeter continuum
spectral energy distribution (SED) based on the ALMA data presented in
this paper and archival centimeter VLA observations from
\citet{2014AJ....147...43K}. This SED contains synchrotron emission,
free-free emission, and thermal emission from dust. Matched
uv-coverage and resolution observations at multiple wavelengths are
required to separate these components and characterize their
properties.

\begin{figure*}
\centering
\includegraphics[width=\textwidth]{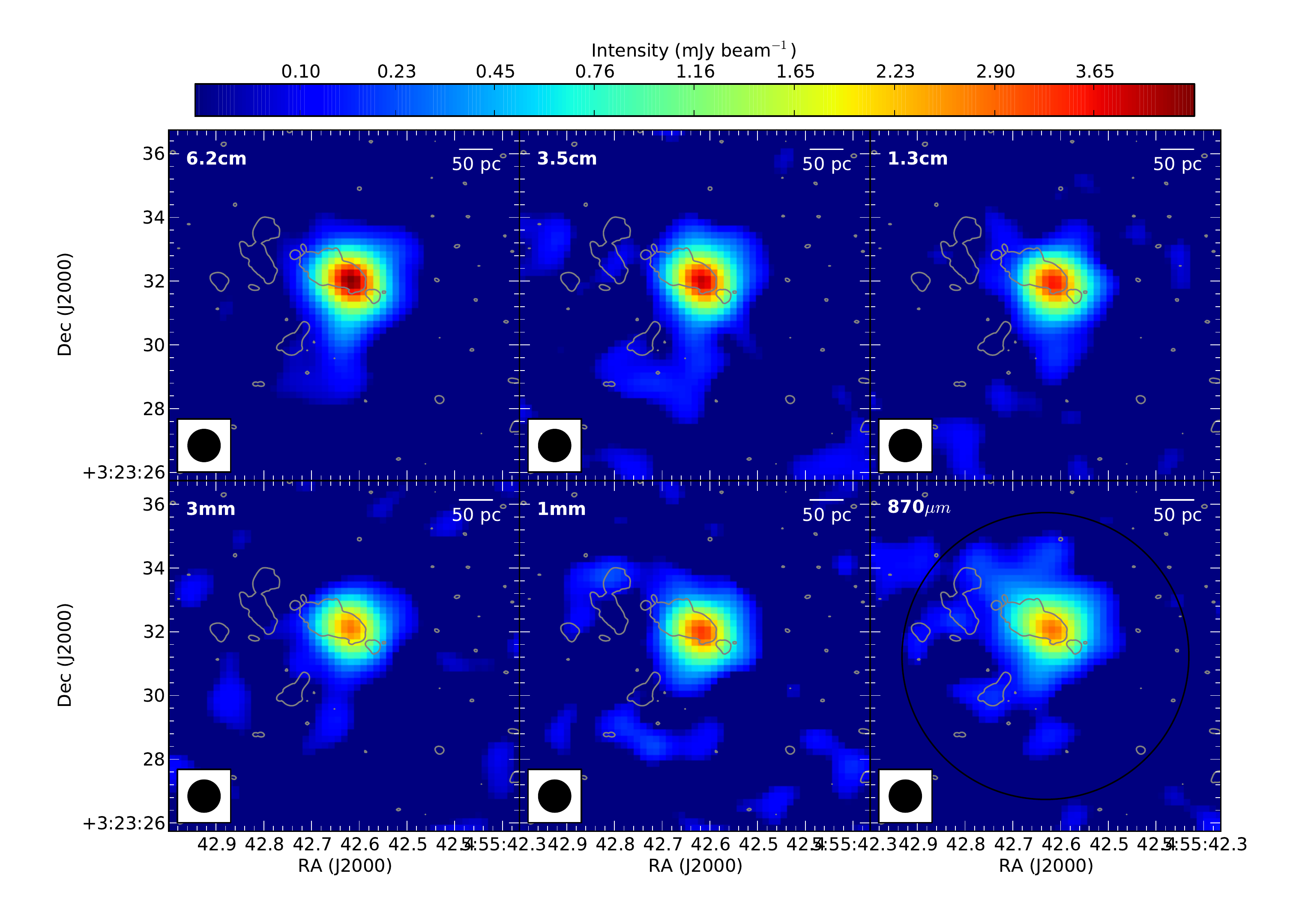}
\caption{Matched resolution and uv-coverage continuum images from
  6.2\,cm to 870\micron\ of \iizw. The contours show the distribution
  of the \cothree\ emission. The wavelength of the image is in the
  upper left, the image scale in the upper right, and the beam in the
  lower left. All six images have the same stretch to bring out the
  changes in extent and brightness of the emission as a function of
  wavelength. The 1\,mm and 3\,mm emission is compact and coincident
  with the 1.3\,cm emission, which is dominated by free-free emission
  \citep{2014AJ....147...43K}. The slightly extended emission at
  6.2\,cm and 870\micron\ is due to the presence of synchrotron and
  dust emission, respectively. The relatively consistent morphology
  and brightness of the continuum emission across all bands suggests
  that the continuum emission in \iizw\ is dominated by free-free
  emission even up to 870\micron.  The large black circle in the
  bottom right panel shows the aperture used to extract the continuum
  flux densities.}
\label{fig:cont_overview}
\end{figure*}

Figure~\ref{fig:cont_overview} shows matched beam and uv-coverage
images for \iizw\ from 6.2\,cm through 870\micron. The 6.2\,cm,
3.5\,cm, and 1.3\,cm data were originally presented in
\citet{2014AJ....147...43K} and have been re-imaged to match the
uv-coverage and beam size of the ALMA observations. See
Section~\ref{sec:data} for details on the imaging process. The 3\,mm
continuum emission is compact and coincident with the free-free
dominated 1.3\,cm emission \citep{2014AJ....147...43K}. The emission
is slightly extended at the highest (1\,mm and 870\micron) and lowest
wavelengths (6.2\,cm) indicating that the presence of emission from
dust and synchrotron emission, respectively. We note that the
morphology of the extended 870\micron\ emission is similar to that of
the \cothree\ emission (overlaid as contours), while the 6.2\,cm
emission extends further south toward the older star-forming region
(SSC-South using the terminology of
\citealp{2014AJ....147...43K}). These images suggest that the
millimeter and submillimeter continuum emission from this galaxy is
dominated by free-free emission and that the dust does not dominate
the continuum emission from \iizw\ even at 870\micron.

\begin{figure}
\centering
\includegraphics[width=\columnwidth]{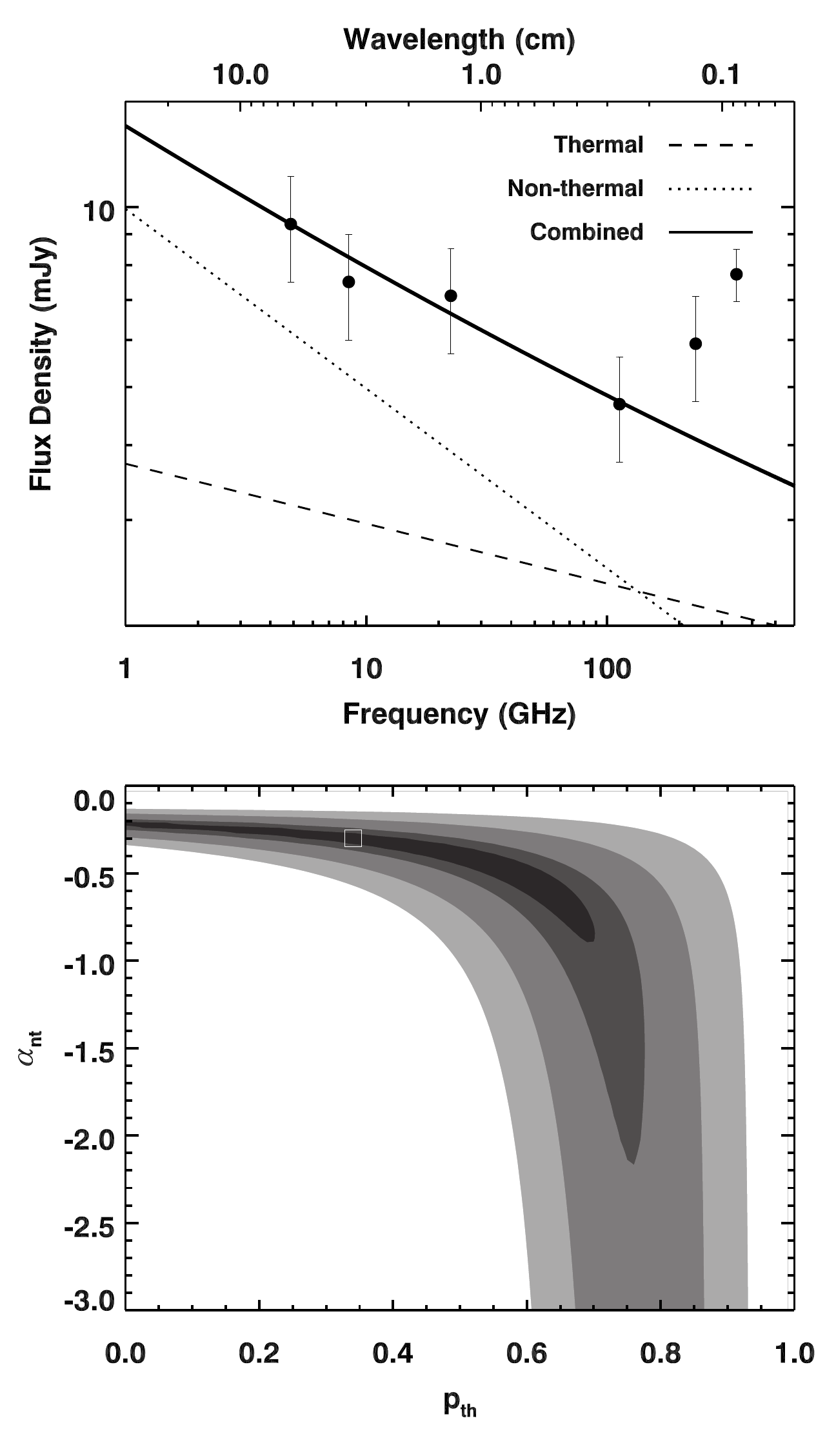}
\caption{{\em Top:} The global continuum spectral energy distribution
  for \iizw\ from centimeter to millimeter wavelengths derived from
  matched uv-coverage and resolution images from the VLA
  \citep{2014AJ....147...43K} and ALMA (this paper). The emission in
  this range consists of three components: synchrotron, which is
  strongest at lower frequencies, free-free, which has a mostly flat
  spectrum, and dust, which is strongest at the highest frequencies in
  this plot. The lines show the best two power law fit to the data
  points less than 230GHz. This plot shows that the synchrotron and
  thermal emission both contribute to the continuum spectral energy
  distribution at frequencies less than 100GHz and that only the 1\,mm
  and 870\micron\ points include a contribution from dust. {\em
    Bottom:} Reduced $\chi^2$ values for fits as a function of thermal
  fraction and non-thermal spectral index. The contours are 1.25, 2,
  5, and 10 times the minimum reduced $\chi^2$ value. Our best fit
  value is shown as a white square.}
\label{fig:cont_sed_nodust}
\end{figure}

To quantify this further, we have extracted the continuum spectral
energy distribution (SED) in a $9\arcsec$ diameter aperture centered
at (5:55:42.63161, 3:23:31.2387). The measured flux densities are
given in Table~\ref{tab:radio_sed}. We fit a simple two power law
model to these values:
\begin{equation}
  \frac{S}{S_0} = p_{th} \left(\frac{\nu}{\nu_0}\right)^{-0.1} 
+ (1 - p_{th} )  \left(\frac{\nu}{\nu_0}\right)^{\alpha_{nt}} 
\end{equation}
where $S$ is the flux density at frequency $\nu$ compared to the flux
density ($S_o$) at a fiducial frequency ($\nu_0$), $p_{th}$ is the
thermal fraction emission, and $\alpha_{nt}$ is the spectral index of
the non-thermal emission.  The aperture was selected to completely
encapsulate the continuum emission at all frequencies. For our fit, we
excluded the 1\,mm and 870\micron\ data since these data points
contain a contribution from dust and set $\nu_0$ to 4.86~GHz. The fit
is shown in Figure~\ref{fig:cont_sed_nodust}. The best fit thermal
fraction is 0.34 with a range of 0 to 0.7 and the best fit non-thermal
spectral index is -0.3 with a range of -0.21 to -0.87. The reduced
$\chi^2$ value for the best fit is 0.18.  This fit shows that, up to
100~GHz, the emission from \iizw\ is a combination of free-free and
non-thermal emission. Only at frequencies greater than 100GHz do we
see any indications of dust emission, and the contribution at those
frequencies is relatively low: even at 870\micron, the dust emission
is only 50\% of the total. We note that the present fit contains a
greater contribution from synchrotron emission than the fit to only
the VLA data points shown in \citet{2014AJ....147...43K}. This
difference is due to the different uv-coverage and resolution of the
images. Here we have emphasized sensitivity to large-scale flux at the
expense of resolution, making the synchrotron emission more pronounced
because it is typically on larger scales. We note that the spectral
index of the synchrotron emission is consistent between the old and
new fits.

\input{iizw40_cont_allbands_nat_uvtaper_1arcsec_v1}

\begin{figure}
\centering
\includegraphics[width=\columnwidth]{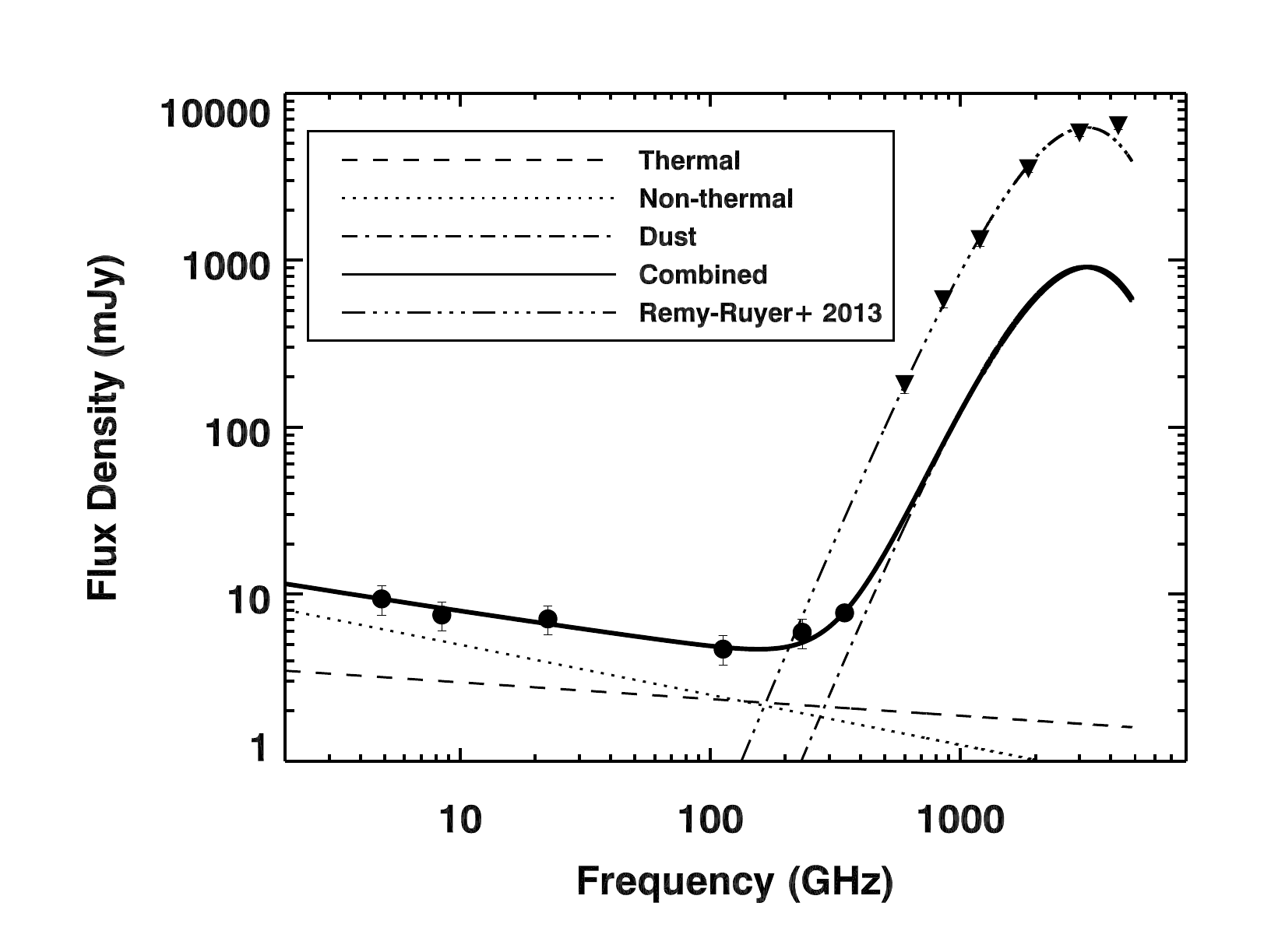}
\caption{The global continuum spectral energy distribution for \iizw\
  from the centimeter (VLA; \citealp{2014AJ....147...43K}),
  millimeter/submillimeter (ALMA), and the far-infrared (Herschel;
  \citealp{2013A&A...557A..95R}).  We include a modified black body
  fit to the 1mm and 870\micron\ points as well as the fit to the
  Herschel data points from \citet{2013A&A...557A..95R} as a
  dot-dot-dash line. The VLA and ALMA points are shown as circles and
  the Herschel points are shown as inverted triangles. For the fit to
  the ALMA data points, the temperature and $\beta$ were fixed to the
  value derived from the Herschel measurements by
  \citet{2013A&A...557A..95R} because this value is largely determined
  by the value of the dust peak traced by Herschel. We find a dust
  mass of $1.9 \times 10^4$ \Msun, which is a factor of 10 lower than
  the dust mass derived using only Herschel photometry by
  \citet{2013A&A...557A..95R}. This result suggests that either a
  large amount of dust emission is being resolved out or that there is
  significant dust emission outside the aperture used for the ALMA
  photometry.}
\label{fig:cont_sed_dust}
\end{figure}

To further explore the dust content of \iizw, we have added the mid-
and far-infrared Herschel photometry and resulting fit from
\citet{2013A&A...557A..95R} to the ALMA and VLA data points to better
capture the peak of the dust emission
(Figure~\ref{fig:cont_sed_dust}). To obtain an estimate of the dust
mass from these points, we attempted to fit a modified black body to
the dust emission:
\begin{equation}
  F_\nu = \frac{M_{dust} \kappa (\lambda_0)}{D^2}
  \left(\frac{\lambda}{\lambda_0}\right)^{-\beta} B_\nu(\lambda,T)
\end{equation}
where $M_{dust}$ is the dust mass, $\kappa(\lambda_0)$ is the dust
absorption coefficient at the reference wavelength $\lambda_0$, $D$ is
distance to the object, $\lambda$ is the observed wavelength, $\beta$
is the dust emissivity index, and $B_\nu(\lambda,T)$ is the Planck
function. We assume a value for $\kappa(\lambda_0)$ of
4.5~${\rm m^2 \, kg^{-1}}$. This value is based on the models of
\citet{2004ApJS..152..211Z} and is the same value used in
\citet{2013A&A...557A..95R}.  All attempts to fit a modified blackbody
through both the ALMA and Herschel data points yielded poor fits. The
failure of these fits is most likely due to either large-scale dust
emission being resolved out by ALMA or to the large difference between
the apertures used for the Herschel and ALMA photometry: 132\arcsec\
versus 9\arcsec.  Although we could increase the aperture size for the
ALMA data, the Herschel aperture is larger than our ALMA field of view
(see Table~\ref{tab:image_summary_cont}).

Given our inability to fit both the ALMA and Herschel points
simultaneously, we derive an estimate of the dust mass using only the
ALMA data by setting $\beta$ and the temperature to the values derived
by \citet{2013A&A...557A..95R}: 1.71 and 33K. We obtain a dust mass of
$(1.8\pm0.2) \times 10^4 \ \Msun$, which is a factor of 10 less than
the dust mass estimated by \citet{2013A&A...557A..95R}:
$1.9 \times 10^5 \ \Msun$. We note that fixing the $\beta$ value only
has a minor effect on the derived mass. Changing $\beta$ from 1.71 to
2.0 changes the mass by less than 10\%.  In addition, since our
observations are on the Raleigh-Jeans tail of the dust distribution,
our derived dust mass is only minimally affected by the assumed
temperature: the mass-weighted temperature would need to have a factor
of two uncertainty to produce a factor of two uncertainty in the mass
.


Armed with a dust mass for \iizw, we can calculate its gas-to-dust
ratio (\gdr), which links the gas and dust content within the
galaxy. The gas-to-dust ratio is used as an alternative way to trace
the molecular gas content of a galaxy
\citep[e.g.,][]{1997A&A...328..471I,2011ApJ...737...12L,2013ApJ...777....5S}. Here
we use archival VLA neutral hydrogen data \citep{1998AJ....116.1186V}
and the molecular gas measurements from
Section~\ref{sec:coh-ratio-molecular} to calculate the gas-to-dust
ratio in \iizw\ and compare it to theoretical estimates of the
gast-to-dust ratio in low metalllicity systems.  

The gas-to-dust ratio is defined as
\begin{equation}
  \gdr = \frac{\aco \LCO + M_{HI} }{M_{dust}}
\end{equation}
where $\aco$ is the \coh\ conversion factor in $\MsunKkmspc$, $\LCO$
is the CO luminosity in $\Kkmspc$, and $M_{HI}$ and $M_{dust}$ are the
neutral hydrogen and dust masses, respectively, in \Msun. Here we are
defining \gdr\ in terms of masses, instead of surface densities, which
requires us to match the apertures used to measure each quantity. In
the same 9\arcsec\ aperture as we used to measure the continuum SED,
we measure an $M_{HI}$ of $6.0\times10^6$~\Msun\ and a \cothree\
luminosity of $2.1 \times 10^5$~\Kkmspc, which corresponds to a
\coone\ luminosity of $3.8 \times 10^5$~\Kkmspc\ using an \rthreeone
value of 0.54. From \S~\ref{sec:coh-ratio-molecular}, we have
estimates from \aco\ that range from 18.1 \MsunKkmspc\ to 150.5
\MsunKkmspc. These values give a range for \gdr\ of 720 to 3500. We
note that, if we are resolving out a significant amount of continuum
emission, our \gdr\ will be overestimates (because we will have
underestimated the dust content). The lower end of our estimated range
range is similar to the \citet{2014A&A...563A..31R} predicted value
(645-660, depending on assumptions about \aco). It is greater than,
but still within an order of magnitude, of the \gdr\ predicted by the
\citet{2011ApJ...737...12L} (330) and the range of values given for
blue compact dwarfs of similar metallicity by
\citet{2014A&A...561A..49H}. We do not see any evidence for the
dichotomy seen in Hunt et al.\ for the \gdr\ ratios for the very low
metallicity galaxies I~Zw~18 and SBS0335-052, although the metallicity
of \iizw\ is ten times higher than in both of those systems.


\subsection{The Molecular Clouds Fueling the Starburst within \iizw}\label{sec:molec-clouds-fuel}

Observations of low metallicity galaxies with the spatial and spectral
resolutions necessary to resolve individual molecular clouds are rare
because of the faint nature of the CO emission in these
galaxies. Previously, only a dozen or so galaxies had bright enough CO
to obtain high resolution observations with sufficient signal to
noise, e.g., Table~1 in \citet{2008ApJ...686..948B}. These galaxies
were typically limited either to very nearby galaxies or to those with
relatively high metallicities (and thus brighter CO). The sensitivity
and resolution of ALMA observations, like those presented here for
\iizw, allow us to measure for the resolved molecular cloud properties
in fainter and more extreme low metallicity systems, and compare them
to the properties of molecular clouds in other galaxies.

Originally motivated by observations in the Milky Way
\citep{1981MNRAS.194..809L,1987ApJ...319..730S}, the so-called
Larson's laws are commonly used to quantify and compare the giant
molecular cloud properties in galaxies.  The first relationship states
that the linewidth of a molecular cloud scales with its size and is
generally referred to as the size-linewidth relationship. The second
relationship shows that the CO luminosity of a molecular cloud is
correlated with its virial mass. This relationship is typically used
to infer the conversion factor between the observed CO luminosity
and the molecular gas mass of a cloud, under the assumptions that
virial equilibrium holds and that the CO traces the full extent of the
molecular gas. The final relationship, which is a consequence of the
first two, states that molecular clouds all have approximately the
same surface density.  These relationships have been shown to be
sensitive to the resolution and sensitivity limits of the data and the
methods used to measure the cloud properties
\citep[e.g.,][]{2011ApJS..197...16W,2013ApJ...779...46H}. However,
they continue to provide the basic framework for understanding giant
molecular cloud populations.

In this section, we identify individual molecular clouds within \iizw,
quantify their properties, and compare these properties to those of
molecular clouds in other environments. Our comparison sample was
selected to span a wide range of cloud properties and environments. It
includes cloud samples from the Milky Way disk
\citep{2009ApJ...699.1092H}, the center of the Milky Way
\citep{2001ApJ...562..348O}, the nearby low metallicity irregular
Large Magellanic Cloud \citep{2011ApJS..197...16W}, the Local Group
dwarf starburst IC 10 \citep{2006ApJ...643..825L}, the major merger
referred to as the Antennae (\citealp{2014ApJ...795..156W}, Leroy et
al.\ in prep), and the nuclear starburst galaxy NGC 253
\citep{2015ApJ...801...25L}. To mitigate the effects of varying
sensitivity and resolution, we have focused on comparison cloud
samples whose original observations have relatively high
($\lesssim 50$pc) resolution and good surface brightness sensitivity
and where the algorithms used to determine the cloud properties were
similar to those used here, i.e., variants on the algorithm in
\citet{2006PASP..118..590R}.

\subsubsection{Measuring the Cloud Properties} \label{sec:meas-cloud-prop}

We used the cprops algorithm \citep{2006PASP..118..590R} to identify
individual molecular clouds within \iizw\ and calculate their
properties. This algorithm identifies significant emission within a
cube using a dilated mask and associates this emission with individual
clouds using a modified watershed algorithm.  The \cothree\ data cube
was used to identify the clouds because it has the highest
signal-to-noise out of all our data sets.  To identify significant
emission, we used a three dimensional noise cube and required the
emission to be greater than 5$\sigma$ in two adjacent channels to
generate an initial mask. All pixels connected to this initial mask
down to a 2$\sigma$ level were then added to the mask. To decompose
the emission into individual clouds, we used a box of 0.35\arcsec\ by
0.35\arcsec\ by 6~\kms; 0.35\arcsec\ is 17pc at the adopted distance of
\iizw. We rejected regions with a contrast less than 1.5$\sigma$ and
areas less than the FWHM of the beam. We note that emission that can
be associated with more than one cloud is not included in the
mask. The measured cloud properties have been extrapolated down to 0K
and corrected for the angular and velocity resolution of the data as
described in \citet{2006PASP..118..590R}.

\begin{figure*}
\centering
\includegraphics[width=\textwidth]{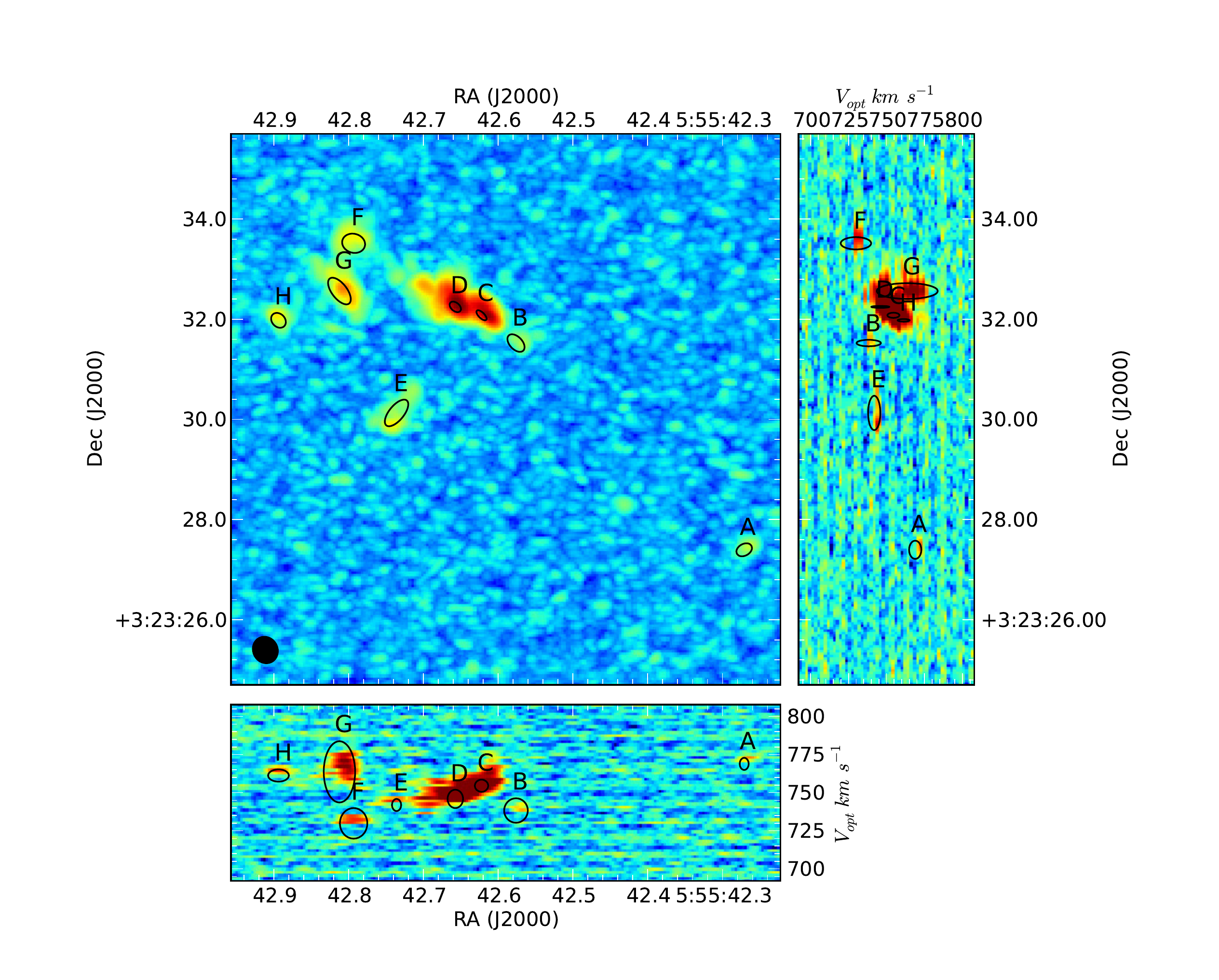}
\caption{Clump assignments as measured using the cprops algorithm
  overlaid on an image of the \cothree\ cube in RA/Dec space (large
  square), RA/velocity space (bottom rectangle), and Dec/Velocity
  space (left rectangle). } \label{fig:cprops_clumps}
\end{figure*}

Our results do not depend strongly on the cube used for the
decomposition, decomposition algorithm, or decomposition
parameters. The cube decompositions for both the \cotwo\ and \cothree\
both identify the brightest clouds.  The main difference between the
two cloud identifications is that the \cothree\ emission spans a
larger continuous area at high signal to noise; the \cotwo\ emission
appears to be moderately clumpier because only the brightest peaks
remain at high signal to noise. An alternative decomposition algorithm
-- clumpfind \citep{1994ApJ...428..693W} -- produces similar results
when applied to our data to those generated by the cprops
algorithm. Finally, changing the decomposition parameters only affects
how many clouds the main molecular ridge is separated into. Depending
on the parameters chosen, the ridge can either be identified as a
single large cloud or up to four different clouds. We have tuned these
parameters to match the number of clumps that can be distinguished by
eye in the position-velocity plot of the \cothree\ emission
(Figure~\ref{fig:cprops_clumps}).

The cloud identifications are shown in Figure~\ref{fig:cprops_clumps}
overlaid on the \cothree\ data and the cloud properties in
Table~\ref{tab:gmc_props}.  Since the clouds are largely Gaussian, the
final cloud sizes and luminosities (and thus their associated
properties like mass, etc) were taken from the Gaussian fit to the
cloud sizes with the beam and channel width deconvolved. For clouds
that were unresolved, we derive an upper limit on the radius of the
cloud by calculating the diameter of a circle that has the same area
as the area at half-maximum covered by the cloud.

Monte Carlo simulations were used to determine the errors on the
measured properties. We added Gaussian noise with the same properties
as the noise cube used to mask the emission to the image cube and
calculated the properties of the resulting clumps. After 100
iterations, we calculated the standard deviation of the resulting
clump properties and used this value as the error. For flux-related
values, we added the errors on the derived flux (10\% at 870\micron\ and
20\% at 1\,mm; see Section~\ref{sec:data}) in quadrature with the
errors derived from the Monte Carlo simulations.

In the literature, the \coone\ transition is the most common tracer of
molecular gas.  To compare our data to the \coone\ data in the
literature, we scale our \cothree\ emission for all clouds, except
cloud G, to the expected \coone\ value using a 10K blackbody, which is
consistent with the line ratios for the entire galaxy (see
Section~\ref{sec:overview-data}). Since cloud G has a significantly
different value of \rthreetwo, we scale its \cothree\ emission to the
expected \coone\ value using a 6K blackbody, which is consistent with
the ratio seen in that cloud (again see
Section~\ref{sec:overview-data}).

\subsubsection{The \coh\ Conversion
  Factor} \label{sec:coh-ratio-molecular}

Since CO is used as the primary tracer of molecular gas beyond the
Local Group, quantifying the relationship between CO emission and the
amount of molecular gas in different environments is key for
understanding the role of molecular gas in star formation throughout
cosmic time. In the Milky Way, this conversion factor is typically 4.3
\MsunKkmspc\ \citep{2013ARA&A..51..207B}. However, it is a factor of
$\sim5$ lower in starburst galaxies like LIRGs and ULIRGs and several
times higher in low metallicity systems
\citep{2013ARA&A..51..207B}. As detailed in the introduction, the high
\aco\ values in the latter systems are due to reduced dust shielding
for the CO \citep[e.g.][]{2008ApJ...686..948B,2011ApJ...737...12L}. In
contrast, the lower values of \aco\ in starburst galaxies are due to
increased CO luminosity due to higher gas temperatures and broader
velocity widths \citep[e.g.,][]{2012MNRAS.421.3127N}.

\begin{figure}
\centering
\includegraphics[width=\columnwidth]{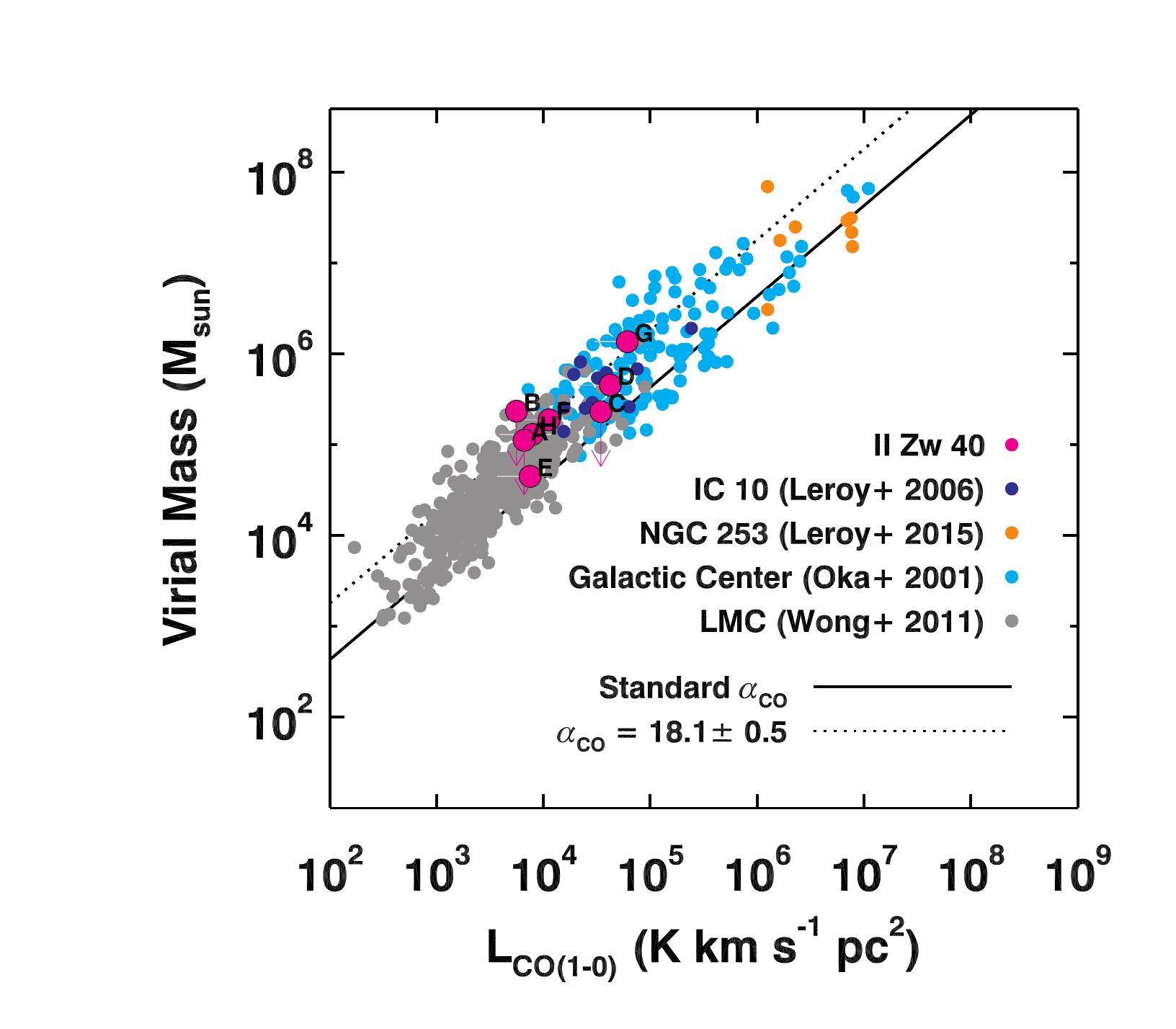}
\caption{The virial mass of the molecular clouds identified in
  \S~\ref{sec:meas-cloud-prop} as a function of their CO
  luminosity. We also include clouds from our comparison samples: the
  Local Group dwarf starburst IC 10 \citep{2006ApJ...643..825L}, the
  nuclear starburst NGC 253 \citep{2015ApJ...801...25L}, the Milky Way
  center \citep{2001ApJ...562..348O}, and the Large Magellanic Cloud
  \citep{2011ApJS..197...16W}. Assuming that the clouds in \iizw\ are
  in virial equilibrium, the relationship between these two quantities
  yields an estimate of the \coh\ conversion factor (\aco). A fit to
  the resolved molecular clouds in this plot suggests that
  $\alpha_{co}$ is approximately four times its value in the Milky
  Way. This conversion factor is similar to the conversion factors
  seen in other low metallicity galaxies, rather than the conversion
  factors seen in more massive starbursting galaxies (LIRGs/ULIRGs).}
\label{fig:LCO_virmass}
\end{figure}

Assuming that the clouds in \iizw\ are in virial equilibrium, we can
derive \aco\ for molecular clouds within \iizw\ by comparing the
virial masses of the clouds with their CO luminosities.  For \iizw, we
derive a value for \aco\ of $18.1\pm0.5$~\MsunKkmspc\ by fitting a
slope to the resolved giant molecular cloud population in
\LCO-$M_{vir}$ space (Figure~\ref{fig:LCO_virmass}). This value is
approximately four times higher than the Milky Way value, but is
consistent with \aco\ values in other dwarf starburst galaxies derived
using resolved CO observations \citep{2008ApJ...686..948B}. Two
exceptions to this trend are Clouds E and C. Cloud E, the faintest
cloud with virial mass estimate in our sample, is a more diffuse
molecular cloud located south of the main star-forming
region. However, given the errors on the associated luminosity, the
\aco\ value for this cloud is broadly consistent with the derived
\aco\ value above. Cloud C is associated with the central
starburst. This cloud has the lowest ratio of virial mass to CO
luminosity, suggesting that its \aco\ value may be more like that of a
starburst galaxy. However, given that we only have an upper limit on
its virial mass, this remains a suggestion rather than a definite
statement.

Our \aco\ estimate is dependent on our assumed value of \rthreeone. If
\rthreeone\ is higher, i.e., the clouds are hotter, then the estimated
\coone\ values will be smaller and the fitted \aco\ factor higher. If
\rthreeone\ is lower, i.e., the clouds are cooler, then the estimated
\coone\ values will be larger and the fitted \aco\ factor lower. We
note that we would require an \rthreeone\ value of 0.12, implying an
extremely cold cloud temperature ($\lesssim 3K$), for the \aco\ value
derived from our data to be equal to the Milky Way value. This value
is clearly at odds with the line ratios seen in \iizw\
(\S~\ref{sec:co-emission-iizw}). We include the statistical errors on
\rthreeone\ (see Section \ref{sec:co-emission-iizw} for a dicussion)
in our \coone\ luminosities in Figure~\ref{fig:LCO_virmass}. Based the
errors for \rthreeone\ ranges, it is unlikely that our derived \aco\
would be off by more than the error in \rthreeone\ ($\sim$~30\%),
although the unresolved points only have upper limits for their virial
masses, and thus could drive the \aco\ factor lower.
 
In low metallicity systems, estimates of the CO-to-H$_2$ conversion
factor derived using virial masses are systematically lower than the
estimates derived using IR observations
\citep[e.g.,][]{2011ApJ...737...12L}. This offset is thought to be
caused by the reduced dust content of low metallicity systems: dust is
the primary shielding mechanism for CO, but H$_2$ molecules are able
to self-shield. At solar metallicity, the H$_2$ and CO emission have
similar physical extents. As the metallicity decreases, the amount of
dust decreases, which decreases the shielding available for CO,
driving it toward the centers of the molecular clouds
\citep{2008ApJ...686..948B}. The molecular gas not traced by CO is
commonly referred to as the ``dark'' molecular gas
\citep{2010ApJ...716.1191W}.

The systematic differences between CO and IR-based \aco\ estimates
suggests that the \aco\ estimate for \iizw\ derived here may
underestimate the true \aco\ value for \iizw. Ideally, we would
compare the CO-based \aco\ estimate for \iizw\ to a measured IR-based
value of \aco\ using far-infrared dust maps and neutral hydrogen
\citep{2011ApJ...737...12L,2013ApJ...777....5S}. However, the spatial
resolutions of the best possible neutral hydrogen and far-infrared
observations are far too low to resolve the central star-forming
region of \iizw: the maximum feasible resolution of neutral hydrogen
observation is $\sim 5\arcsec$, while far-infrared observations from
Herschel have resolutions ranging from $\sim 5\arcsec$ to
$\sim 30\arcsec$.  Although the 870\micron\ continuum data presented
here could be used to provide an estimate for the dust emission, the
extended dust emission is only seen in the heavily tapered, low
resolution image, not in an image with uv-coverage and resolution
matched to that of the \cothree\ observations.

We can use the total CO luminosity and derived dust mass along with an
estimate of the gas-to-dust ratio (\gdr) to estimate a value of \aco\
for the central star-forming region of
\iizw. \citet{2013A&A...557A..95R} provide an empirical estimate of
the \gdr\ as a function of metallicity. For the metallicity of \iizw\
(8.09; \citealp{2000ApJ...531..776G}), the estimated \gdr\ is between
645-660. Here we adopt a mid-range value of 650. The total estimated
\coone\ luminosity in this region (adopting the typical value of
\rthreeone\ in \S~\ref{sec:meas-cloud-prop}) is
$3.8 \times 10^5 \ \Kkmspc$, the dust mass for the central
star-forming region is $1.4 \times 10^4 \ \Msun$ (see
\S~\ref{sec:cont-emiss-iizw}), and the neutral hydrogen mass for the
same region is $6.0\times 10^6 \ \Msun$.  These values give an
estimate of \aco\ of 14\MsunKkmspc, which is consistent with the
CO-based \aco\ estimate. However, this value is a lower limit, given
that we may be underestimating the dust mass because we are resolving
out continuum emission.

Finally, we can use the simple photodissociation model developed by
\citet{2010ApJ...716.1191W} to estimate \aco. In this model, \aco\
depends strongly on the column density of the gas, which for resolved
observations depends on the metallicity and integrated CO line
intensity. Thus, with our resolved observations, we can independently
derive an estimate for \aco\ based only on these two readily observed
quantities.  In the Appendix, we derive an
expression for the ratio of \aco\ at a metallicity $Z$ to \aco\ at
solar metallicity as a function of CO line intensity $W_{CO}$ and
metallicity ($Z$). Taking values appropriate for \iizw\ ($Z=0.2$ and
$W_{CO(1-0)}$ between 0.85 and 2.6 \Kkms), we find that, according to
this model, the \aco\ for \iizw\ should be 15 to 35 times higher than
the Milky Way value, lending support to the idea that the \aco\ value
derived using virial mass estimates is indeed a lower limit on the
\aco\ value in this galaxy. However, there is significant scatter in
the CO line intensity values, suggesting that the assumption of
constant CO surface brightness leading to constant molecular gas
surface densities may not be valid. For \iizw, this leads to a factor
of two scatter in the estimated \aco value. We will explore the
assumption of constant CO surface density in detail in a later section
(\S~\ref{sec:cloud-surf-brightn}).

Our resolved observations provide the rare opportunity to compare our
cloud-based estimates of \aco\ to model predictions of global \aco\
values in low metallicity environments. \citet{2012MNRAS.421.3127N}
uses the results of \citet{2010ApJ...716.1191W} to model populations
of molecular clouds within galaxies. From their simulations, they
derive a relationship between the surface brightness of CO and the
metallicity of a galaxy. With our cloud-scale resolution, we are in a
unique position to test the veracity of these models at low
metallicity. Using their Equation 7, the metallicity of \iizw\, and
the average \coone\ surface brightness of its clouds from
Figure~\ref{fig:LCO_size} (1.8~\Kkms), we derive an estimate of \aco\
of 17.5, which is close to value we derive from the virial mass
estimate. However, we note that the value derived here is in the low
CO surface brightness limit where, according to
\citet{2012MNRAS.421.3127N}, the properties of the molecular clouds
depend less on environment. Given the observational evidence for
larger \aco\ values in low-metallicity galaxies, we suggest that the
\aco\ values derived from these models may be underestimates for low
metallicity systems.

The preceding discussion of \aco\ estimates has ignored any
time-variable effects on the \aco\ value, which may be essential for
starbursting systems like \iizw. Bursts of star formation like those
found in \iizw\ may increase the dichotomy between the dust-derived
and virial-derived \aco\ estimates by increasing CO
photodissociation. In pre-starburst systems, these \aco\ estimates may
be closer together due to the reduced flux of ionizing photons.


\subsubsection{The Molecular Star Formation
  Efficiency} \label{sec:molecular-gas-star}

With our estimates for \aco\ in hand, we can now derive a value for
the molecular star formation efficiency within \iizw, which
characterizes how well this galaxy is able to turn molecular gas into
stars.\footnote{We use the extragalactic definition for the molecular
  star formation efficiency: SFE=SFR/$M_{H_2}$.} For the entire
central region, \LCOthree\ is $2\times10^5 \ \Kkmspc$, which
corresponds to an \LCOone\ of $3.8\times10^5 \ \Kkmspc$ using the
appropriate value for \rthreeone\ (see \S~\ref{sec:co-emission-iizw}).
To obtain an upper limit on the star formation efficiency, we use our
largest estimate for \aco: 150.5~\MsunKkmspc. The star formation rate
in the central region of \iizw\ is 0.34~$\Msun \, yr^{-1}$
\citep{2014AJ....147...43K}. These values yield a molecular star
formation efficiency for \iizw\ of
$6 \times 10^{-9} \, {\rm yr}^{-1}$, which is approximately 10 times
higher than the average star formation efficiencies found in nearby
spirals \citep{2008AJ....136.2782L}.

If we use the virial-based \aco\ estimate, we derive an even higher
molecular star formation efficiency:
$4 \times 10^{-8} \, {\rm yr}^{-1}$. This value is two orders of
magnitude higher than in normal spirals and is similar to the star
formation efficiencies seen in starburst galaxies
\citep{2010ApJ...714L.118D}.  However, by using the virial-derived
\aco\ measurement, we are only including the molecular gas emission
from deep inside the molecular cloud rather than averaging the
molecular gas over kpc regions as in \citet{2008AJ....136.2782L} and
\citet{2010ApJ...714L.118D}.  Therefore, the molecular star formation
efficiency calculated using a virial-based \aco\ estimate may be
expected to be higher because it includes only the densest molecular
gas regions where stars are more likely to form, while the values
given in \citet{2008AJ....136.2782L} and \citet{2010ApJ...714L.118D}
average over many of these such regions and also include diffuse
molecular gas.

The molecular star formation efficiency values derived above do not
take into account the star formation history of \iizw\ and the effect
of its young massive clusters on their surrounding molecular gas.  In
more massive galaxies, the star formation rate is roughly constant
with time over kpc-sized regions or we would not see the observed
level of agreement between different star formation rate tracers
\citep{2011ApJ...737...67M,2012ApJ...761...97M,2012AJ....144....3L}. A
continuous star formation rate ensures a steady conversion of gas into
stars, allowing the present star formation in these systems to be
linked to the remaining gas supply. However, in the case of
interacting systems like \iizw, their star formation varies with both
position and time and their bursts of star formation dissociate the
leftover molecular gas. These time dependent effects mean that one
cannot directly link the present day star formation with the past
molecular gas supply.

We suggest that the high molecular star formation efficiency seen in
\iizw\ is due to its rapidly changing, merger-driven star formation
history, not an intrinsically high molecular star formation
efficiency. For example, in \iizw, the massive cluster associated the
molecular gas (SSC-N) is young \citep[$\lesssim$
5Myr;][]{2014AJ....147...43K}. The cluster immediately to the south
(SSC-S) is much older (9.5~Myr) and shows no associated molecular
gas. We can also see clear signs that SSC-N is destroying the
surrounding molecular gas. As we saw in \S~\ref{sec:co-emission-iizw},
the \rthreeone\ ratios are elevated here, indicating hotter gas. Higher
resolution observations of the ionized gas within cloud C show
filamentary structures, suggesting that ionizing radiation from SSC-N
is destroying its molecular envelope \citep{2014AJ....147...43K}.
Therefore, understanding the observed molecular star formation
efficiencies in galaxies like \iizw\ requires careful modeling of the
star formation histories in these systems and the effects of the young
massive clusters on the surrounding interstellar medium.




\subsubsection{CO Surface Brightnesses and Mass Surface
  Densities} \label{sec:cloud-surf-brightn}

As discussed in \S\ref{sec:molec-clouds-fuel}, there is increasing
evidence that the third Larson's relation -- molecular clouds have
constant mass surface densities -- is the result of the sensitivity
and resolution limits of earlier surveys
\citep{2011ApJS..197...16W,2013ApJ...779...46H}. However, there do
appear to be systematic offsets in the surface density of clouds
between galaxies that have been attributed to differences in pressure
\citep{2013ApJ...779...46H}. In this section, we explore whether the
molecular clouds within \iizw\ have constant surface brightnesses and
thus constant mass surface densities and compare these quantities to
the surface brightnesses and surface densities in our comparison
sample.

\begin{figure*}
  \plottwo{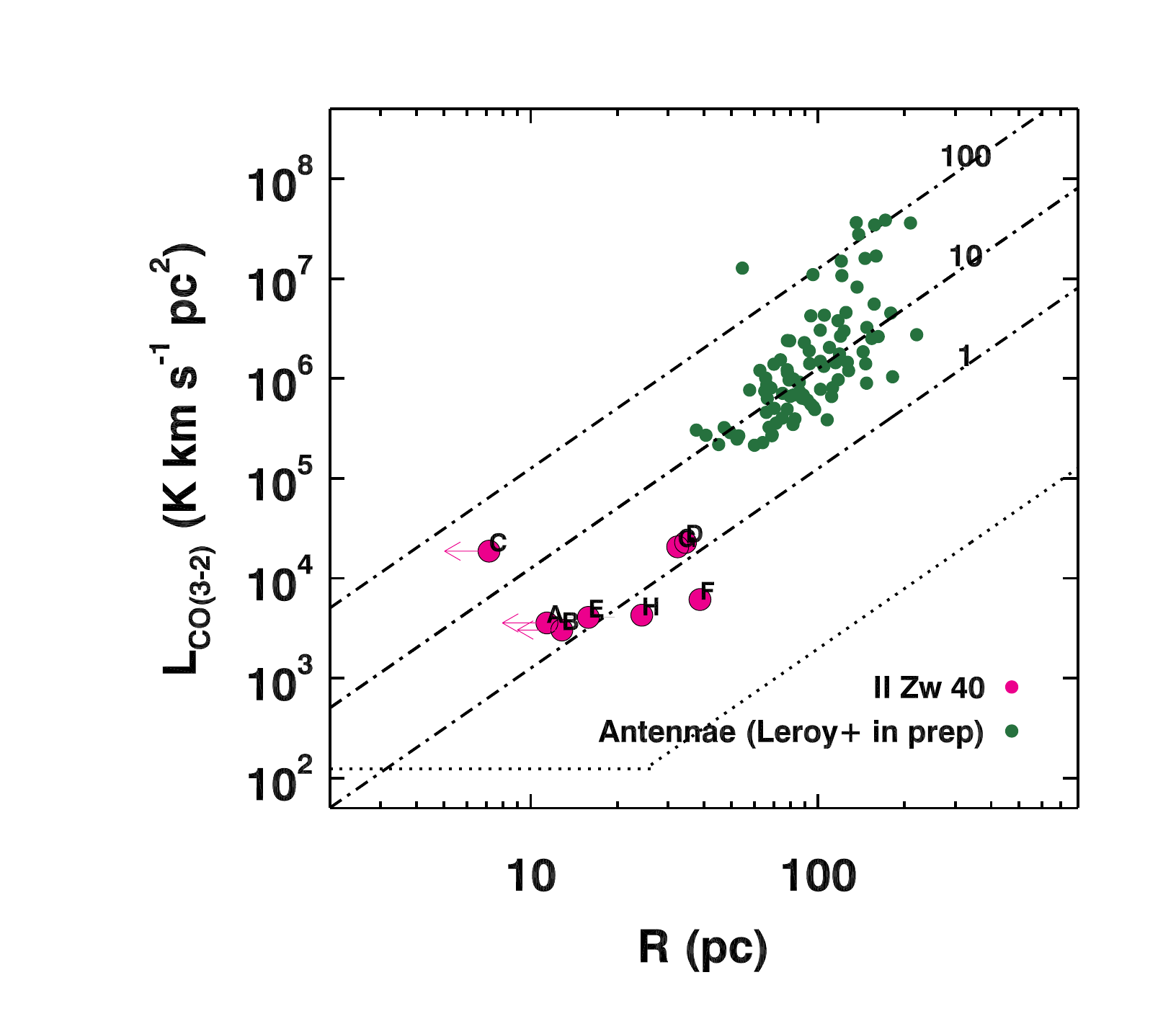}{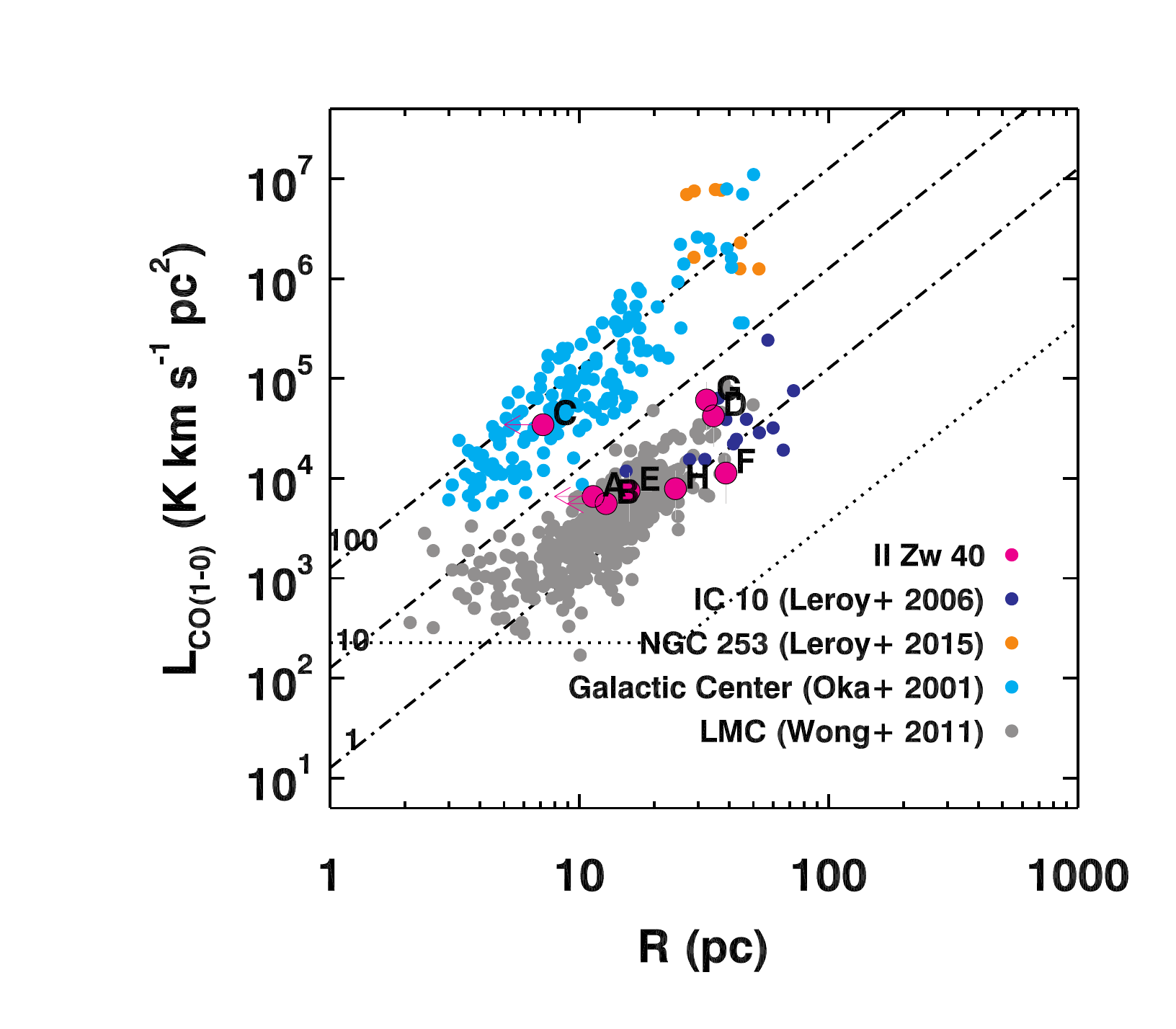}
  \caption{\LCOthree\ (left) and \LCOone\ (right) in \iizw\ and the
    comparison sample as a function of cloud radius. The dash-dotted
    lines are lines of constant surface brightness (1, 10, 100 \Kkms)
    and the dotted line indicates the sensitivity and resolution
    limits of the \iizw\ data. In general, the clouds in \iizw\ have
    lower CO surface brightnesses than the clouds seen in higher
    metallicity galaxies like the Antennae, NGC 253, and the center of
    our own Galaxy. The clouds in \iizw\ have similar surface
    brightness to the clumps in the LMC and the Local Group dwarf
    starburst IC 10. The one exception is cloud C, which higher CO
    surface brightness comparable to the CO surface brightnesses in
    the Galactic Center, NGC 253, and the Antennae. This cloud is also
    the only cloud in \iizw\ with active star formation.}
\label{fig:LCO_size}
\end{figure*}

Figure~\ref{fig:LCO_size} compares CO luminosity as function of cloud
size for both \iizw\ and the comparison sample. The left panel
includes the measured \cothree\ values for both the Antennae and
\iizw\, while the right panel has the estimated \coone\ values for
\iizw\ and the measured \coone\ values from galaxies in the comparison
sample. Lines of constant surface brightness are shown as dot-dashed
lines and the sensitivity and resolution limit of the \iizw\ data is
shown as a dotted line.

These plots show that, for the most part, the clouds in \iizw\ have
lower CO surface brightnesses than clouds in the Galactic Center, NGC
253, and the Antennae. These surface brightnesses are comparable to
clouds found in the LMC and IC 10. Therefore, the globally low CO
luminosity in \iizw\ is due to low overall CO surface brightnesses,
not a few high surface brightness clouds with very low filling
factors. If \iizw\ is representative of dwarf starburst galaxies as a
whole, then we can expect that they will also have low overall CO
surface brightnesses. The exception to the low-surface brightness
trend is Cloud C: it has a comparable surface brightness to clouds in
the Antennae, the Galactic Center, and NGC 253. This cloud is also the
only cloud in \iizw\ with active star formation.

\begin{figure}
\includegraphics[width=\columnwidth]{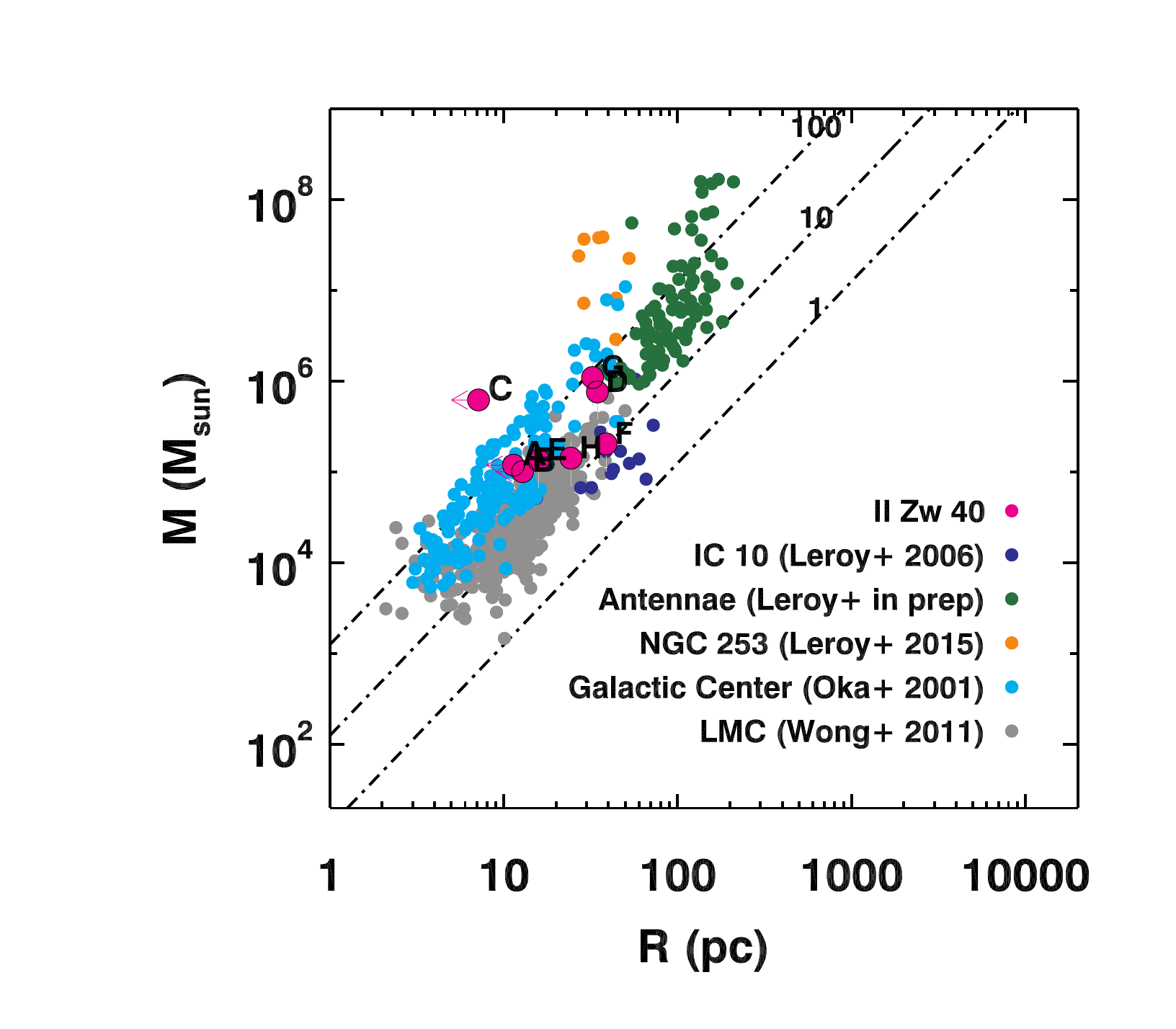}
\caption{CO-derived mass as a function of radius for \iizw\ and the
  comparison sample. The dashed-dotted lines indicated lines of
  constant mass surface density (100, 10, 1 \Msunpc). When the proper
  \aco\ conversion factor is used, the clouds in \iizw\ have similar
  mass surface densities to those in the Galactic Center, the
  Antennae, and NGC 253 and on the upper end of those in IC 10 and the
  LMC. Therefore, the low CO surface brightnesses in \iizw\ are due to
  the lower \aco\ value in this galaxy, not clouds with intrinsically
  low mass surface densities.}
\label{fig:mass_size}
\end{figure}

Figure~\ref{fig:LCO_size} also shows that the molecular clouds within
a particular galaxy tend to have similar CO surface brightness, i.e.,
the clouds fall along a diagonal line in \LCO\ versus radius plots.
For our data, however, any systematic variations in the CO surface
brightness could also be due to differences in \aco, not just
differences in the mass surface density of clouds, To account for the
difference \aco\ values, we plot the CO-derived cloud mass versus
radius for \iizw\ and the comparison sample as a function of size in
Figure~\ref{fig:mass_size}. This plot shows that the CO-derived mass
surface densities of molecular clouds within \iizw\ are on average
higher than those in the LMC or IC 10 and are comparable to the
CO-derived mass surface densities of clouds within the Antennae and
the Galactic Center. Based on these results, the low CO luminosities
seen in \iizw\ are the result of higher \aco\ factors rather than low
surface density molecular gas, showing that the clouds within \iizw\
have normal mass surface densities compared to clouds in other
galaxies.

Systematic uncertainities with the values of \aco\ and \rthreeone\
will impact the location of the \iizw\ clouds on these plots. Smaller
values of \rthreeone, which imply colder molecular gas, will move the
clouds to higher \LCO\ values in the right panel of
Figure~\ref{fig:LCO_size}. For the mass versus radius plots, the
\rthreeone\ and \aco\ affect the result in opposite directions:
increasing \rthreeone\ leads to lower \LCO\ values and thus lower mass
surface densities, while increasing \aco\ leads to higher mass surface
densities. Given that the values we derive for \rthreeone\ and \aco\
are consistent with values derived in similar galaxies (see
\S~\ref{sec:co-emission-iizw} and \S~\ref{sec:coh-ratio-molecular}),
it is relatively unlikely that we have major systematic errors in our
derivations of \aco\ and \rthreeone.

\begin{figure*}
\plottwo{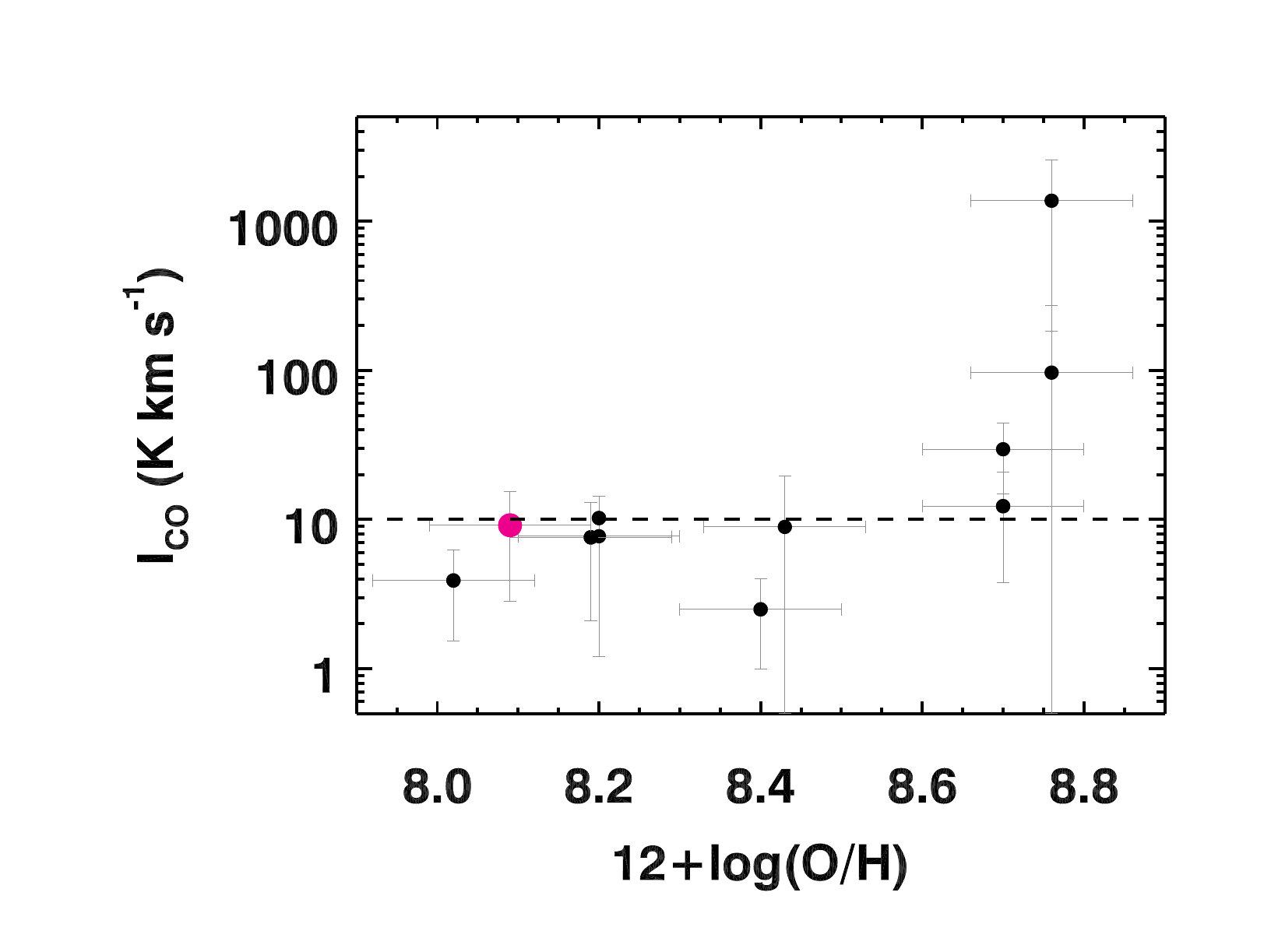}{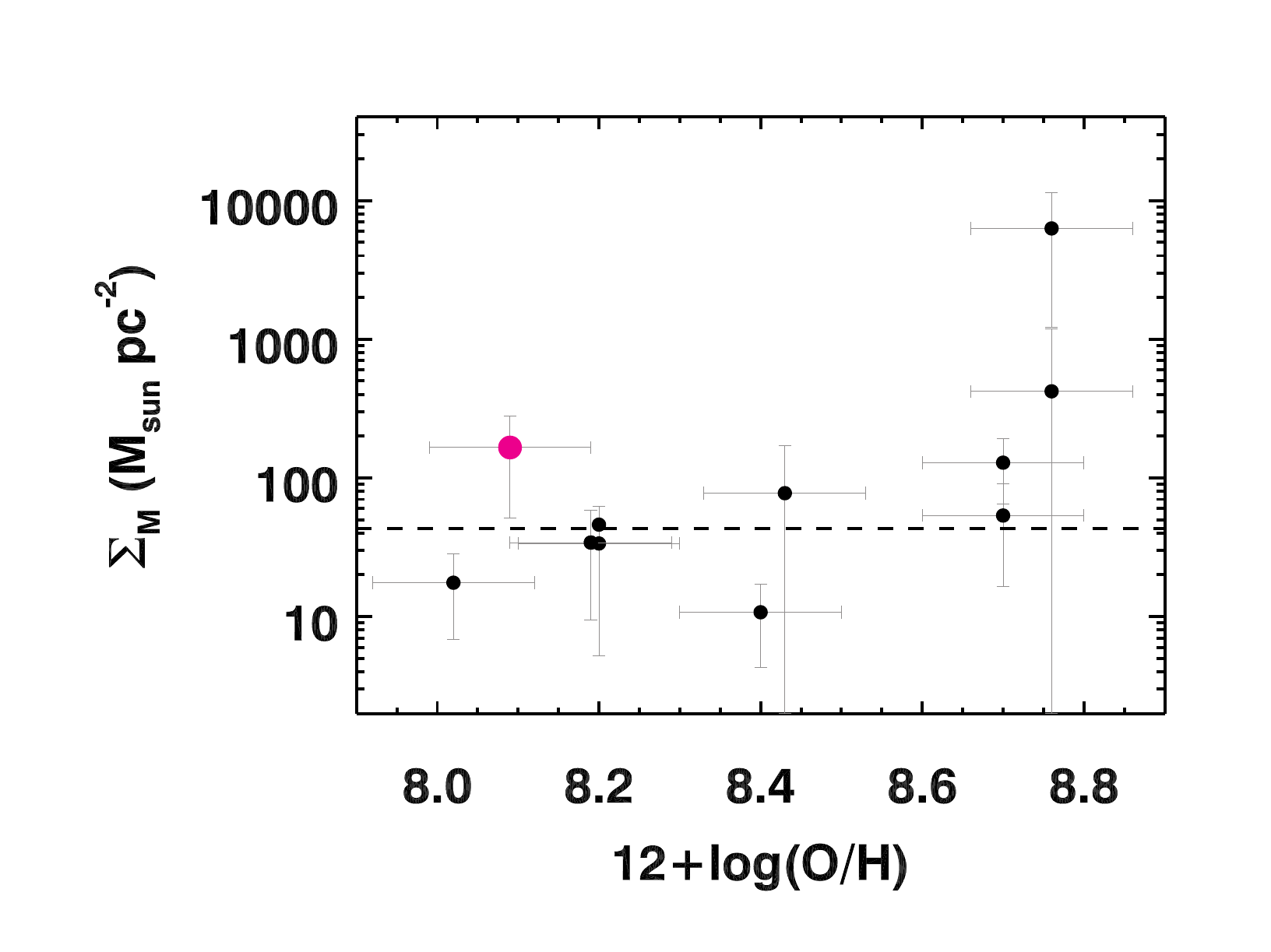}
\caption{CO surface brightness (left) and mass surface density (right)
  of clouds as a function of metallicity for \iizw\ (magenta point)
  and galaxies with CO observations at comparable spatial resolutions
  ($\sim30$pc). The CO surface brightnesses for galaxies with
  metallicities less than 12+log(O/H)=8.6 are less than or equal to
  10~\Kkms (dashed line). However, when the appropriate \aco\ values
  are applied to the data, the molecular surface densities in two of
  our low metallicity galaxies (\iizw\ and the LMC) exceed the
  expected 10~\Kkms\ mass surface density for a Galactic \aco\
  value. Therefore, the low CO surface brightnesses found in low
  metallicity systems does not necessarily imply that these systems
  have low molecular gas mass surface densities.}
\label{fig:surface_plots}
\end{figure*}

To explore the trend of mass surface density with metallicity further,
we plot the CO surface brightness as a function of metallicity in the
left panel of Figure~\ref{fig:surface_plots}.  For this plot, we have
selected comparison galaxies with CO observations at comparable
spatial resolution ($\sim 30$pc) with a range of metallicities
including the Milky Way \citep{2009ApJ...699.1092H}, the LMC
\citep{2011ApJS..197...16W}, IC 10 \citep{2006ApJ...643..825L}, NGC
1569, NGC 205, and SMC \citep{2008ApJ...686..948B}, M31 (Scruba et
al.\ in prep), M33 \citep{2007ApJ...661..830R}, the Antennae
\citep[][; Leroy et al.\ in prep]{2014ApJ...795..156W}, and NGC 253
\citep{2015ApJ...801...25L}. As expected, we find that the CO surface
brightness for galaxies with metallicities less than 12+log(O/H)=8.6
is less than or equal to 10~\Kkms. However, when the proper \aco\
values are applied to the data, we find that at least two of the
galaxies (II Zw 40 and the LMC) have larger mass surface densities
than expected if we applied a Galactic \aco\ value to a CO surface
brightness limit of 10\Kkms. This result shows that the low CO surface
brightnesses seen in low metallicity galaxies do not necessarily mean
that the molecular gas surface densities in these galaxies are low.

\subsubsection{The Size-Linewidth Relationship
} \label{sec:size-linew-relat}


The last fundamental molecular cloud relationship we will investigate
is the size-linewidth relationship, which is thought to arise from the
turbulent cascade of energy within a molecular cloud. The location of
clouds within this plot reflects the combined effects of virial
equilibrium, turbulence, and external pressure. For viralized clouds,
the normalization of this relationship is set by cloud surface
density. As we saw in \S~\ref{sec:cloud-surf-brightn}, the surface
brightnesses of clouds in the Milky Way
\citep{1987ApJ...319..730S,2009ApJ...699.1092H} and in high resolution
studies of nearby galaxies
\citep{2011ApJS..197...16W,2008ApJ...686..948B} are similar and thus
their clouds fall along the same size-linewidth relationship. Higher
surface densities clouds like those seen in the nuclear starburst NGC
253 \citep{2015ApJ...801...25L} and in the Antennae \citep[][Leroy et
al.\ in prep]{2014ApJ...795..156W} have higher normalization factors
due to their high gas surface densities. Finally, clouds can be
elevated beyond the nominal normalization factor derived from their
surface densities by higher external pressures, such as those found in
the center of the Milky Way \citep{2001ApJ...562..348O}.

\begin{figure}
\centering
\includegraphics[width=\columnwidth]{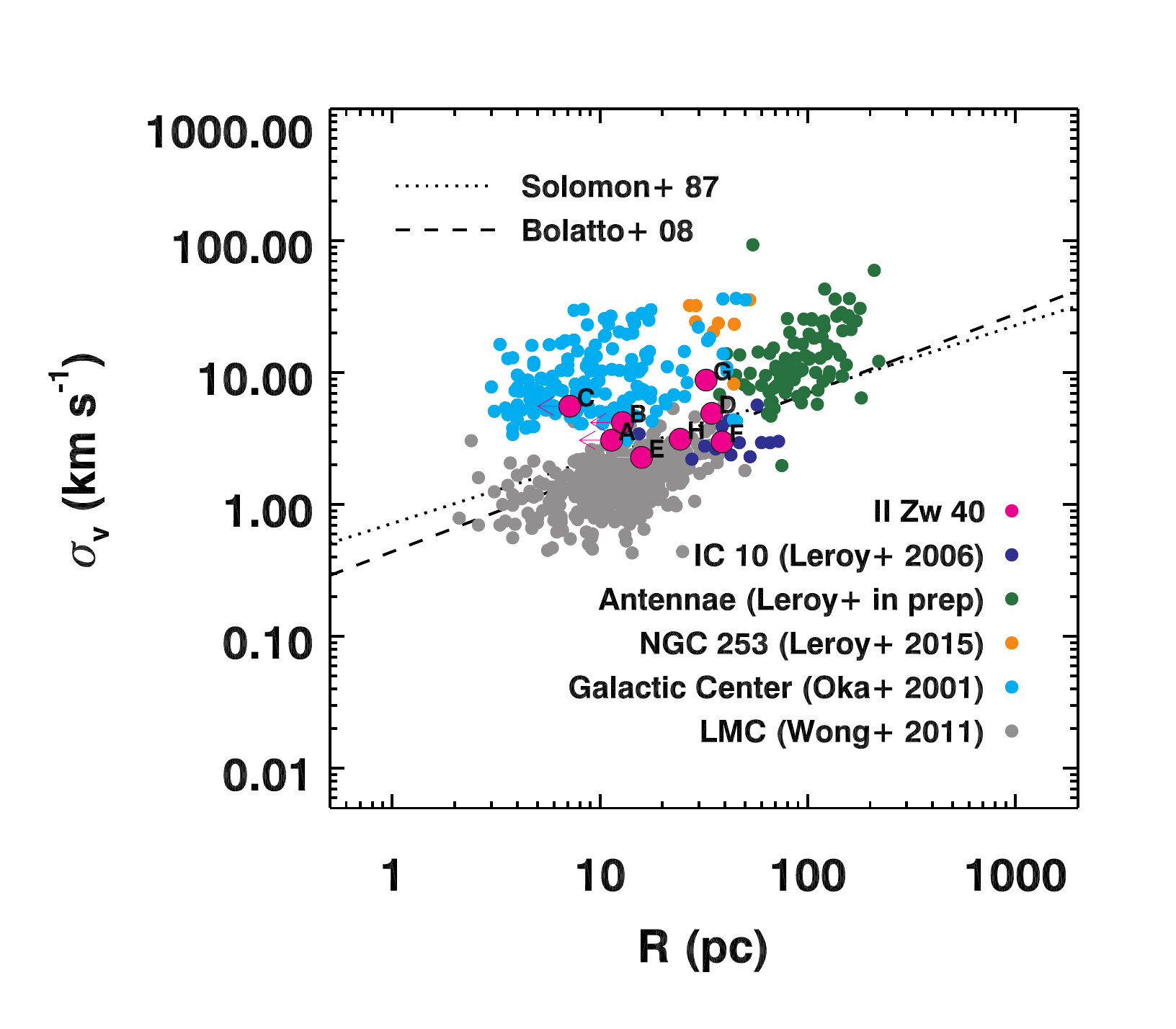}
\caption{The size-linewidth relationship for molecular clouds within
  \iizw\ and our comparison galaxies. The clouds in \iizw\ are shown
  as magenta circles outlined in black and labeled. The points show
  different comparison samples of clouds including: the Large
  Magellanic Cloud \citep[gray points;][]{2011ApJS..197...16W}, the
  Local Group dwarf starburst IC 10 \citep[purple
  points;][]{2006ApJ...643..825L}, the massive late-stage merger
  referred to as the Antennae (dark green points;
  \citealp{2014ApJ...795..156W}, Leroy et al., in prep), the nuclear
  starburst NGC 253 \citep[orange points;][]{2015ApJ...801...25L}, and
  the center of the Milky Way \citep[cyan
  points;][]{2001ApJ...562..348O}. The lines show the fiducial
  size-linewidth relationships from \citet{1987ApJ...319..730S}
  (dotted) and \citet{2008ApJ...686..948B} (dashed). The clouds within
  \iizw\ lie above the fiducial size-linewidth relationship seen in
  the Milky Way and the relationship seen in the LMC and have similar
  velocities for their size to clouds in the Galactic Center, the
  Antennae, and the dwarf starburst IC 10. This trend suggests that
  either external pressure or turbulence play a key role in driving
  the linewidths of the clouds within
  \iizw.}\label{fig:size_linewidth}
\end{figure}

Figure~\ref{fig:size_linewidth} compares the sizes and linewidths of
clouds within \iizw\ to those in our comparison samples. We find that
the molecular clouds within \iizw\ have larger than average linewidths
for their size compared to clouds in the Milky Way, suggesting that
external pressure and/or high gas surface densities play a significant
role in shaping the molecular clouds within this galaxy. We note that
IC 10, one of the few other dwarf starbursts with measured molecular
cloud properties, displays similar properties to \iizw, but that
clouds in the LMC generally have smaller linewidths compared to those
in \iizw.  Deep neutral hydrogen observations suggest that IC 10 is
either an advanced merger or has an infalling gas cloud
\citep{2014AJ....148..130A}.  The molecular clouds in this galaxy also
lie largely on the edges of super-shells produced by stellar winds and
supernovae generated by its on-going burst of star formation
\citep{2006ApJ...643..825L}.

\begin{figure*}
\centering
\includegraphics[width=\textwidth]{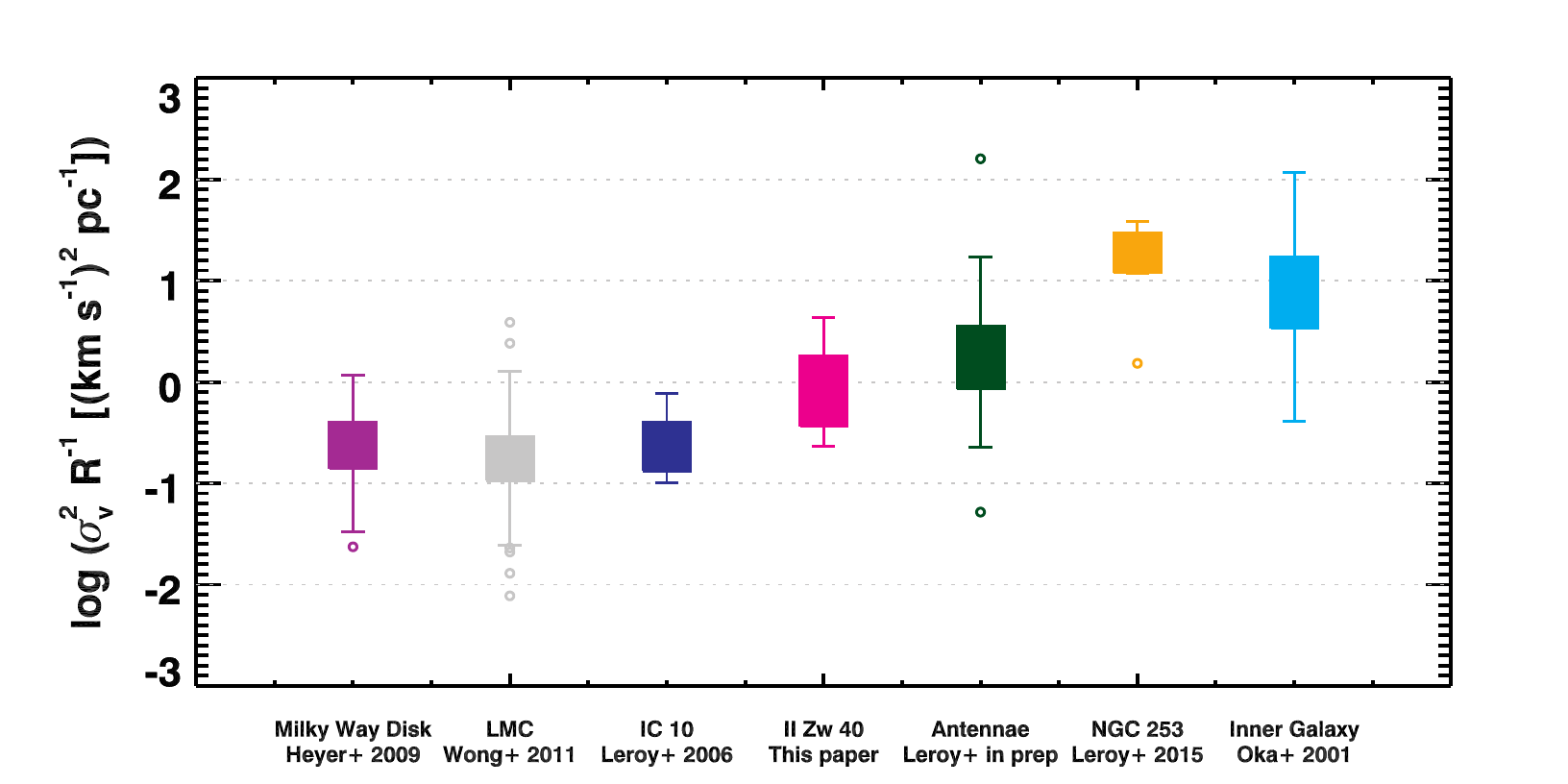}
\caption{Distribution of the coefficient of the size-linewidth
  relationship squared ($\sigma_V^2/R$), which is proportional to the
  GMC surface density in virial equilibrium, for \iizw\ and comparison
  sample. The lower and upper ends of the box demarcate the first and
  fourth quartile. The whiskers at the tops of the box extend to the
  minimum and maximum of the distribution or 1.5 times the mininum or
  maximum. Outliers are shown as circles. Although all samples have a
  wide range of values, the size-linewidth coefficients in \iizw\ are
  most similar to those seen in the Antennae. Like \iizw, the Antennae
  are undergoing a major merger between two galaxies, although for the
  Antennae the progenitor galaxies are more massive than the dwarf
  progenitors involved in the \iizw\ merger. The size-linewidth
  coefficients for the nuclear starburst NGC 253 and the center of the
  Milky Way are, in general, larger than the coefficients for either
  the Antennae or \iizw. The distribution of size-linewidth points in
  IC 10, a dwarf starburst galaxy in the Local Group, is similar,
  although not quite as high as that for \iizw. IC 10 shows signs of
  large-scale gas shocks. The neutral hydrogen distribution of IC 10
  is consistent with either late-stage merger or an infalling gas
  cloud \citep{2014AJ....148..130A} and the molecular gas lies on the
  edges of the super-shells created by the on-going burst of star
  formation in this galaxy \citep{2006ApJ...643..825L}.}
\label{fig:c_boxplot}
\end{figure*}

To further distinguish between the different samples, we compare the
distribution of the square of the coefficient of the size-linewidth
relationship ($C^2 = \sigma_v^2/R$), which is proportional to the
molecular gas surface density if the cloud is in virial equilibrium,
in the different samples (Figure~\ref{fig:c_boxplot}). This plot makes
clear that the \iizw\ clouds have the largest overlap with the clouds
seen in the Antennae, although they do also overlap the upper end of
the Milky Way molecular clouds and the lower end of the molecular
clouds in NGC 253 and the Inner Galaxy. Again, the clouds in \iizw\
and the nearby dwarf starburst galaxy IC 10 have similar values. This
result suggests that the clouds in \iizw\ have higher surface
densities than found in the disk of the Milky Way (if they are in
virial equilibrium).

\begin{figure*}
\centering
\includegraphics[width=\textwidth]{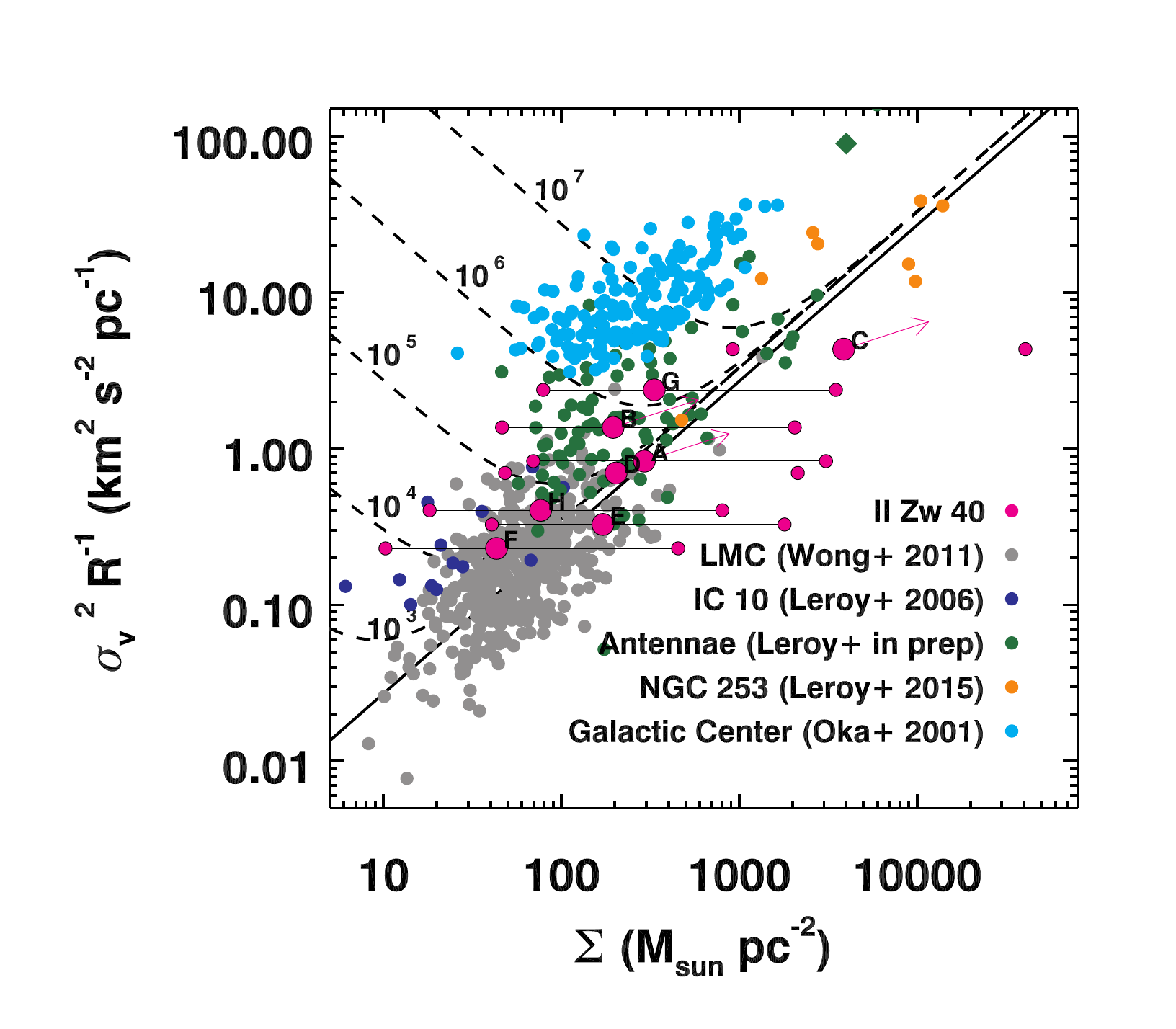}
\caption{The coefficient of the size-linewidth relationship squared
  ($\sigma_V^2/R$) as a function of molecular gas surface density
  derived from CO observations for molecular clouds within \iizw\ and
  our sample of comparison galaxies. Clouds on the solid line are in
  virial equilibrium, while clouds above the line require external
  pressures to maintain equilibrium.  The dotted lines indicate the
  required external pressures in units of $\rm{K \, cm^{-3}}$. The
  clouds in \iizw\ are shown as magenta circles outlined in black. The
  large magenta circles use our CO-based \aco\ measurement (see
  \S~\ref{sec:coh-ratio-molecular}), while the small circles on the
  left side of the diagram use a Milky Way estimate (4.3\MsunKkmspc)
  and those on the right side of the diagram use our dust-derived
  \aco\ estimate (see \ref{sec:coh-ratio-molecular}). The points show
  different comparison samples of clouds including: the Large
  Magellanic Cloud \citep[gray points;][]{2011ApJS..197...16W}, the
  Local Group dwarf starburst IC 10 \citep[purple
  points;][]{2006ApJ...643..825L}, the massive late-stage merger
  referred to as the Antennae (dark green points; Leroy et al., in
  prep) with the candidate super star cluster progenitor referred to
  as the Firecracker \citep{2015ApJ...806...35J} shown as a dark green
  diamond, the nuclear starburst NGC 253 \citep[orange
  points;][]{2015ApJ...801...25L}, and the center of the Milky Way
  \citep[cyan points;][]{2001ApJ...562..348O}.  The \iizw\ clouds will
  lie on the virial equilibrium line in this diagram because we have
  assumed virial equilibrium to derive the \aco\ value. However, we
  argue that the CO-based \aco\ value used here is approximately
  correct because it agrees with \aco\ values determined in other
  systems with similar metallicities. Adopting a Milky Way \aco\
  factor would require that the clouds be confined by external
  pressures ranging from $10^4 \ \rm{K \, cm^{-3}}$ to
  $10^6 \ \rm{K \, cm^{-3}}$, while adopting a dust-based \aco\
  estimate would imply that all of our clouds are sub-virial and
  collapsing. Thus, the large linewidths in \iizw\ appear to be the
  result of the high gas surface densities due to \iizw's on-going
  merger and not large external pressures.}
\label{fig:c_sd_alphascale}
\end{figure*}

To investigate whether high external pressures may be playing a role
in \iizw, we compare the size-linewidth coefficient squared to the
surface density of the molecular gas determined using CO in
Figure~\ref{fig:c_sd_alphascale}.  Clouds in virial equilibrium will
fall along the solid line in this diagram, while clouds requiring high
external pressures will fall above the line. The dashed lines indicate
the required external pressure to maintain virial equilibrium.
However, since we have assumed virial equlibrium to determine \aco\
(see \ref{sec:coh-ratio-molecular}), our clouds will fall along the
virial equilibrium line by design.

To show the effects of different \aco\ assumptions, we include in
Figure~\ref{fig:c_sd_alphascale} points using our CO-based \aco\
measurements (large magenta points) as well as a Milky Way \aco\ value
and the dust-based \aco\ estimate derived in
\ref{sec:coh-ratio-molecular} (small magenta points to the left and
right of the virial equilibrium line, respectively). Using a Milky Way
\aco\ value would imply that our clouds require external pressures
ranging over two orders of magnitude from $10^4 \ \rm{K \, cm^{-3}}$
to $10^6 \ \rm{K \, cm^{-3}}$, while a dust-based \aco\ estimate would
imply that all of our clouds are sub-virial and collapsing.  Although
these effects could arise in strong turbulent flows, we argue that the
more likely origin of the large linewidths with \iizw\ is high surface
densities generated by the large scale gas shocks driven by its
on-going merger since our CO-derived \aco\ factor is consistent with
CO-based measurements of \aco\ in other galaxies with similar
metallicity (see \S~\ref{sec:coh-ratio-molecular}). Given this, we do
not find any strong evidence that the clouds within \iizw\ require
high external pressures to maintain their large linewidths.

The two clouds with the greatest deviations from the size-linewidth
relationship -- clouds C and G -- occupy interesting regions of this
diagram. Cloud C lies below the virial equilibrium line even with
using our adopted \aco\ value. This cloud appears to have already
collapsed and formed multiple young massive clusters. High resolution
radio continuum images find at least three embedded sources with
ionizing photon fluxes comparable to 30 Doradus -- the closest example
of a starburst -- associated with cloud C
\citep{2014AJ....147...43K}. The optical cluster near cloud C has an
age of less than 5Myr, which supports the idea that cloud C recently
collapsed.  Cloud G, which also has line ratios consistent with colder
CO, would require the highest external pressures
($10^6 \ \rm{K \, cm^{-3}}$) to maintain virial equilibrium,
suggesting that it could be a progenitor of a super star cluster
\citep{2015ApJ...806...35J}.

A related key parameter for quantifying the clouds within \iizw\ is
the turbulent Mach number \citep{2015ApJ...801...25L}:
\begin{equation}
\mathcal{M}=\sqrt{3}\sigma_{1d} / c_s
\end{equation}
where $\sigma_{1d}$ is the one dimensional velocity dispersion and
$c_s$ is the sound speed. We adopt a value of 0.2~\kms\ for $c_s$,
which assumes an isothermal molecular gas with a temperature of
10K. We note that the gas temperature may be higher closer to the
star-forming region. The value used here reflects the average
temperature of the molecular clouds throughout the central region.
The median Mach number of clouds in \iizw\ is $\sim36$ with values
ranging from 20 to 76. Comparing these values with those for giant
molecular clouds in the Milky Way and in NGC 253 (see Table 4 in
\citealp{2015ApJ...801...25L}), we find that the median Mach number
within \iizw\ is higher than that found in molecular clouds in the
Milky Way disk (11), but below that of the clouds in the nuclear
starburst galaxy NGC 253 (85). The Mach number within \iizw\ is
consistent with the increased Mach numbers necessary to produce
virialized clouds.

Overall, these results suggest that the large-scale gas shocks due to
\iizw's on-going merger are driving high molecular gas surface
densities in \iizw, leading to elevated linewidths. We caution that
\iizw\ is clearly a rapidly evolving system. It is possible that some
(or all) of the molecular gas structures we see here are transient,
unbound features that will not collapse to form stars. However, our
estimates of \aco\ are consistent with other virial equilibrium based
estimates of this value at low metallicity (see
\S~\ref{sec:coh-ratio-molecular} for a discussion), which suggests
that the \iizw\ clouds identified here are most likely bound objects.

\section{Summary and Conclusions} \label{sec:conclusions}

In this paper, we present high-resolution ($\sim24$~pc) ALMA
observations of \cothree\ and \cotwo\ emission and 3\,mm, 1\,mm, and
870\micron\ continuum emission in the prototypical blue compact dwarf
galaxy \iizw. Our goal is to characterize the molecular gas content
within this unique, low metallicity and high star formation rate
surface density environment and compare it to the molecular clouds
properties seen in other environments to gain insight into the
fundamental physical processes driving star formation in galaxies.

Due to the combination of relatively bright line emission and
excellent surface brightness sensitivity, our highest signal-to-noise
CO detection is \cothree\ in ALMA Band 7. The distribution of the CO
emission tracing the molecular clouds and various star formation
tracers is consistent with a picture of merger-driven star formation
within \iizw.  The CO emission within this galaxy is clumpy and
distributed throughout the central star-forming region of \iizw\ and
its kinematics are similar, but not identical to, that of the galaxy's
neutral hydrogen reservoir. Only one of the molecular clouds traced by
the CO emission has associated star formation, suggesting that the
other clouds, in particular the relatively cold cloud G, will serve as
future sites for star formation or are transient molecular features.

The centimeter through submillimeter continuum spectral energy
distribution of \iizw\ is dominated by free-free and synchrotron
emission. Dust emission only begins to play a role at 1\,mm and is
only 50\% of the total emission at 870\micron. We derive a total dust
mass of $1.4 \times 10^4 \ \Msun$ and a gas-to-dust ratio of between
770 and 3500. The lower end of this range is consistent with other
estimates based on far-infrared data. The gas-to-dust ratio derived
here may be an overestimate if we have resolved out significant
continuum emission.

Using the high spatial resolution provided by ALMA, we have used the
\cothree\ data to measure the giant molecular cloud sizes, line
widths, and luminosities within \iizw\ and compared their properties
with populations of clouds in other fiducial star-forming environments
including the Galactic Center \citep{2001ApJ...562..348O}, the Large
Magellanic Cloud \citep{2011ApJS..197...16W}, the Local Group low
metallicity irregular IC 10 \citep{2006ApJ...643..825L}, the nuclear
starburst NGC 253 \citep{2015ApJ...801...25L}, and the merging
Antennae system \citep[][Leroy et al.\ in prep]{2014ApJ...795..156W}.

Using our CO and continuum data, we find that the \coh\ conversion
factor in \iizw\ is at least four times higher than in the Milky Way
(18.1 \MsunKkmspc) and may be as much as 35 times higher (150
\MsunKkmspc). The lower limit is based on an average virial equilibrium
estimate for the resolved clouds in \iizw. This value is comparable to
CO-based virial equilibrium measurements in other low metallicity
galaxies. The upper limit is based on a new method that uses simple
photodissociation models by \citet{2010ApJ...716.1191W} and the
resolved line intensity to uniquely predict \aco. We note that the
\coh\ conversion factors based on the numerical models of
\citet{2012MNRAS.421.3127N} may underestimate the conversion factor,
especially in low metallicity systems.

We use the derived conversion factor to estimate the molecular star
formation efficiency of \iizw. We find that even using our largest
estimate of the \coh\ conversion factor the molecular star formation
efficiency of \iizw\ is still larger than the typical value by a
factor of 10. This high value could be due to truly high molecular
star formation efficiencies in these systems, or more likely,
time-dependent evolutionary effects due to \iizw's on-going merger.
We also find that the low CO surface brightnesses in \iizw\ and other
low metallicity galaxies from the literature do not necessarily imply
that these systems have low molecular gas surface densities. The low
CO surface brightness in these galaxies appears to be due to reduced
dust shielding for CO, not to intrinsically low surface density
molecular gas.

The properties of the clouds within \iizw\ reveal that the large-scale
gas shocks produced by its on-going merger and subsequent star
formation may play a key role in shaping the molecular clouds.  On
average, the clouds in \iizw\ lie above the Milky Way size-linewidth
relationship and have higher size linewidth coefficients. Their
size-linewidths coefficients, which are proportional to the virial
mass surface densities, are comparable to those found the Antennae,
which is a late-stage merger, and IC 10, whose neutral hydrogen
distribution is consistent with either an advanced merger or an
infalling gas cloud and whose molecular gas lies on the edges of
large-scale shells due to on-going star formation. They are lower than
those in the nuclear starburst NGC 253 or the Galactic Center.
Increased surface densities leading to higher linewidths appear to the
origin of the elevated size-linewidth coefficients in \iizw, rather
than large external pressures. We note that this result does depend on
the adopted \aco\ value.

Using \iizw\ as a template, we can extend our results to make several
inferences about properties of molecular gas in blue compact
dwarfs. First, the properties of the molecular gas in \iizw\ are
driven by the large-scale gas shocks induced by its on-going merger,
not by its metallicity. The low metallicity of this galaxy only
affects the observability of the most commonly used molecular gas
tracer CO.  \iizw\ is not a unique case; many blue compact dwarf
galaxies appear to be the result of mergers or interactions
\citep{2012ApJ...748L..24M,2014MNRAS.445.1694L,2015ApJ...805....2S}. Indeed,
unresolved, single-dish \cothree\ and \coone\ observations of a sample
of eight dwarf starburst galaxies have also found that the line ratios 
and derived densities and temperatures of the molecular gas in their
sample have more in common with that of starburst galaxies than of low
metallicity galaxies \citep{2001AJ....121..740M}.

Second, as \iizw\ shows, the star formation histories of blue compact
dwarfs are complex and vary rapidly with time, unlike larger spiral
galaxies which more or less continuously form stars at a low rate. In
other words, blue compact dwarfs are more like firecrackers than the
slow-burning bonfires of spiral galaxies. Therefore, comparing present
day gas content to current star formation to derive molecular star
formation efficiencies and other related quantities in blue compact
dwarfs is misleading at best. The high molecular gas star formation
efficiencies seen here in \iizw\ and claimed for other galaxies
\citep{2015Natur.519..331T} are likely the result of the massive burst
of star formation -- which establishes a source as a blue compact
dwarf galaxy -- using up a majority of the molecular gas in a
galaxy. Building up larger samples of molecular gas in dwarf galaxies
in different evolutionary stages -- now possible with the power of
ALMA -- is key to understanding the properties of their molecular gas
and star formation in these sources.

Based on our results, we suggest two avenues for future studies of the
molecular gas in these unique systems. First, CO detection experiments
for low-metallicity galaxies like \iizw\ should focus on the \cotwo\
or \cothree\ transitions, which are significantly brighter than the
ground state transition. In particular for ALMA, Band 7 observations
of \cothree\ should have the highest signal-to-noise, assuming that
other blue compact dwarf galaxies have similar CO excitations.

Second, although the sensitivity and resolution of ALMA allow us to
extend detailed CO studies to galaxies like \iizw, the use of other
molecular gas tracers in these galaxies needs to be explored.  One
commonly used tracer of molecular gas is dust continuum
emission. Although this is a powerful tracer, as we have demonstrated
here, the dust emission in these galaxies is faint even at 870\micron\
and blended with free-free emission. Higher resolution
($\sim 1\arcsec$) far-infared and neutral hydrogen observations would
be another way to probe the molecular gas content of these galaxies,
but require the development of a high resolution far-infrared
telescope and something like the Square Kilometer Array (SKA). Another
promising tracer might be CI \citep{2015arXiv150901939G} or CII
\citep{2015arXiv151104689G}, which may be prevalent throughout
molecular cloud envelope, but may also extend beyond the cloud.

\acknowledgments The authors would like to thank Dr.\ Liese van Zee
for generously sharing her neutral hydrogen data and the referee for
his or her thoughtful comments. A.A.K. thanks the Collective Agency
(Portland, OR) for hosting her while a portion of this work was
written. A.A.K. would also like to thank John Hibbard, Mark Krumholz,
and Jennifer Donovan Meyer for helpful
conversations. K.E.J. gratefully acknowledges support provided in part
by NSF through award 1413231. The National Radio Astronomy Observatory
is a facility of the National Science Foundation operated under
cooperative agreement by Associated Universities, Inc. This paper
makes use of the following ALMA data: ADS/JAO.ALMA\#
2012.1.00984.S. ALMA is a partnership of ESO (representing its member
states), NSF (USA) and NINS (Japan), together with NRC (Canada) and
NSC and ASIAA (Taiwan), in cooperation with the Republic of Chile. The
Joint ALMA Observatory is operated by ESO, AUI/NRAO and NAOJ. This
research also made use of APLpy, an open-source plotting package for
Python hosted at http://aplpy.github.com


\appendix

\section{Derivation of Predicted \coh\ Conversion Factor as a Function
  of Metallicity and Integrated CO Line Intensity} \label{sec:deriv-pred-coh}

In the \citet{2010ApJ...716.1191W} models of the dark molecular gas,
the estimated \coh\ conversion factor depends strongly on the average
extinction through the giant molecular cloud, which is turn depends
on both the metallicity of the system and the column density of the
giant molecular cloud. For extragalactic observations that do not
resolve molecular clouds, the column density must be assumed and is
commonly set to values similar to those found in Milky Way clouds. For
resolved observations like those presented in this paper, the observed
CO brightness of the cloud adds an additional constraint
\begin{equation} \label{eq:n22}
\bar{N}_{22} = \Xco(Z) W_{CO} / (1 \times 10^{22} cm^{-2})
\end{equation}
where $W_{CO}$ has units of \Kkms and $\Xco$ has units of ${\rm
  cm}^{-2} \, ({\rm K} \,\kms)^{-1}$.  Thus, in the resolved case, the
estimated \coh\ conversion factor depends both on the metallicity, the
CO line intensities, and itself. In this appendix, we solve for the
estimated \coh\ conversion factor as a function of metallicity and CO line
intensity.

Following \citet{2010ApJ...716.1191W}, \citet{2013ARA&A..51..207B}
derive the following relationship between the mean extinction through
a cloud and value of \aco\ relative to the value at solar metallicity
\begin{equation} \label{eq:aco_vs_av}
\frac{\Xco\ (Z')}{\Xco\ (Z'=1)} = \exp{ \left[ \left( \frac{4.0 \Delta
    A_V}{\bar{A}_{V,MW}}\right) \left( \frac{1-Z'}{Z'} \right) \right]}
\end{equation}
where $Z'$ is the metallicity of the galaxy relative to solar,
$\Xco(Z')$ is the value of \Xco\ at metallicity $Z'$, $\Xco(Z'=1)$ is
the value of \Xco\ at solar metallicity
\citep[$2\times10^{-2} \ {\rm cm}^{-2} \, ({\rm K}
\,\kms)^{-1}$;][]{2013ARA&A..51..207B},
$\Delta A_v$ difference in extinction between the CO-dark layer of gas
and the CO-emitting layer, and $\bar{A}_{V,MW}$ is the mean extinction
through the giant molecular cloud.\footnote{The original expression in
  \citet{2013ARA&A..51..207B} is in terms of $X_{CO}$. However, since
  \aco\ scales linearly with $X_{CO}$, this equation also gives the
  ratio of $\aco(Z')/\aco(Z'=1)$.}  The mean extinction through the
giant molecular cloud is given by the expression
\begin{equation} \label{eq:av_n22}
\bar{A}_V = 5.26 \, \delta_{DGR}' \, \bar{N}_{22}
\end{equation}
where $\delta_{DGR}'$ is the dust-to-gas ratio relative to the local
Galactic value and $\bar{N}_{22}$ is the mean hydrogen column density
of the cloud in units of $10^{22} \, \rm{cm}^{-2}$.  For
$\bar{A}_{V,MW}$, the quantity $\delta_{DGR}'$ equals one.
Substituting Equations~(\ref{eq:n22}) and (\ref{eq:av_n22}) into
Equation (\ref{eq:aco_vs_av}), we obtain
\begin{equation}
\frac{\Xco\ (Z')}{\Xco\ (Z'=1)} = \exp{ \left[
    \left(\frac{\Xco(Z'=1)}{\Xco(Z')} \right) \left( \frac{ 7.604
        \times 10^{21} \Delta
    A_V}{W_{CO} \Xco(Z'=1)}\right) \left( \frac{1-Z'}{Z'} \right) \right]},
\end{equation}
which we rearrange to obtain
\begin{equation}\label{eq:xco_final}
  \frac{\Xco(Z')}{\Xco(Z'=1)} \ln{\left( \frac{\Xco(Z')}{\Xco(Z'=1)}
    \right)} = \left( \frac{ 7.604 \times 10^{21} \Delta    A_V}{W_{CO} \Xco(Z'=1)}\right) \left( \frac{1-Z'}{Z'} \right).
\end{equation}

\begin{figure}
\centering
\includegraphics[width=\columnwidth]{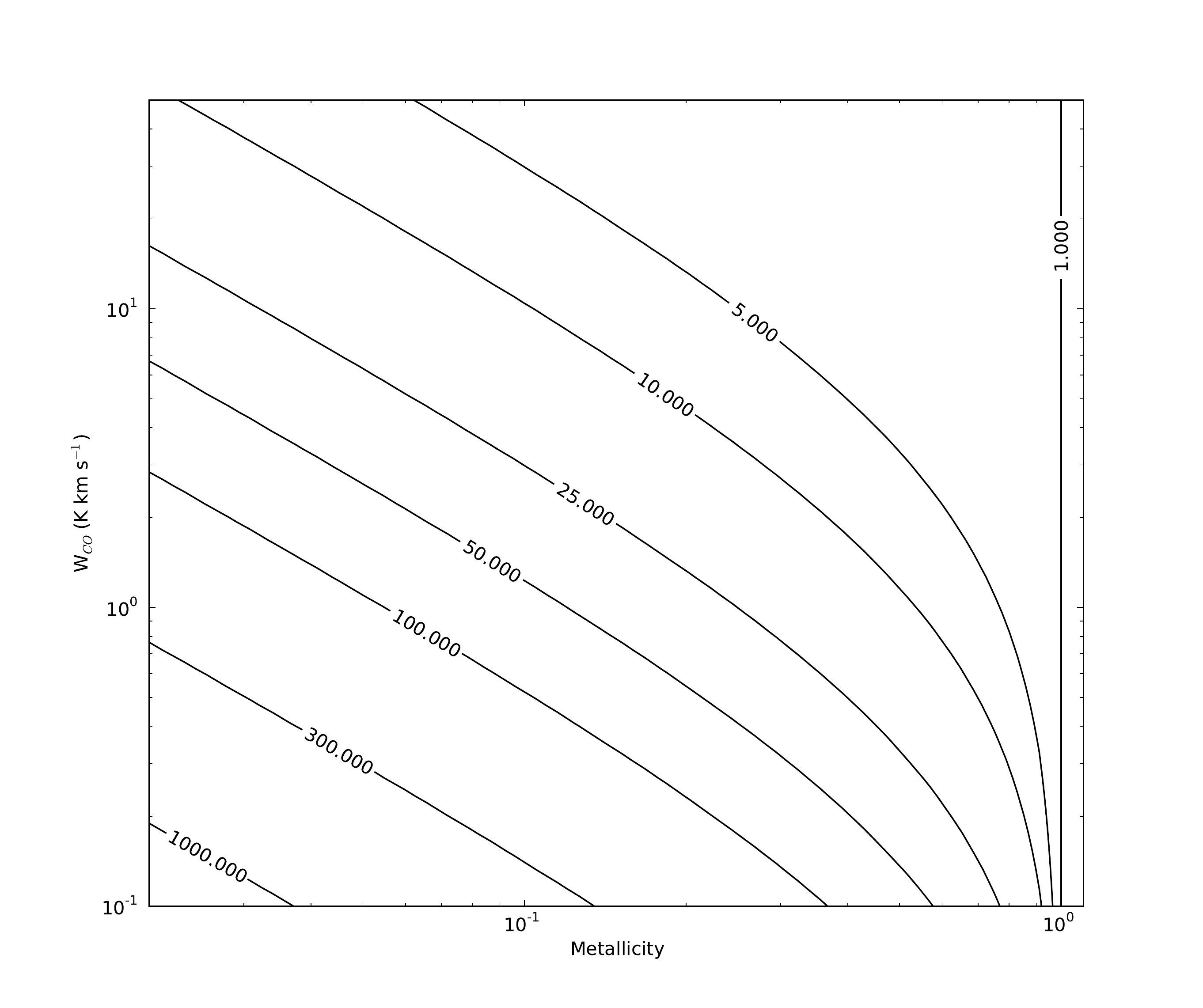}
\caption{The ratio $\Xco(Z')/\Xco(Z'=1)$ as a function of integrated
  line intensity ($W_{CO}$) and metallicity relative to solar
  ($Z'$). This calculation takes into account the strong dependence of
  the fraction of dark molecular gas on the column density of the
  molecular gas.}
\label{fig:alpharatio}
\end{figure}

We use Equation~(\ref{eq:xco_final}) to calculate how the value of
\Xco\ changes with the $W_{CO}$ and $Z'$. The term $\Delta A_V$ has
only a weak dependence on metallicity, so we adopt the mean value for
$\Delta A_V$ from \citep{2010ApJ...716.1191W}: 0.7. The solution to
this equation is of the form
\begin{equation}
x = \exp{(W(c))}
\end{equation}
where $W(c)$ is the Lambert W function, $x$ is $\Xco(Z')/\Xco(Z'=1)$,
and $c$ is the right hand size of Equation~(\ref{eq:xco_final}). The
value for the \Xco\ ratio is plotted in Figure~\ref{fig:alpharatio} as
a function of integrated line intensity ($W_{CO}$) and metallicity
($Z'$).

\input{iizw40_gmc_props_final_v1_astroph}

\end{document}

%% file: observation_summary.tex
\begin{deluxetable*}{lccc}
\tablecolumns{4}
\tablewidth{0pt}
\tabletypesize{\scriptsize}
\tablecaption{Observation Summary\label{tab:obs_summary}}
\tablehead{
\colhead{} & \colhead{3\,mm} & \colhead{1\,mm} & \colhead{870\micron\ } }
\startdata
Date                                 &  2014 August 30  & 2014 June 14   &   2013 Oct 09    \\
Number of Antennas                   &   35             &  33            &      19              \\
Mean Wind Speed (\ms)                &    9             &   7            &      6               \\
Mean Precipitable Water Vapor (mm)   &    3.38          &   0.51         &      0.42            \\
Flux Calibrator                      &  J0510+180       &  J0510+180     &    J0510+180         \\
Bandpass Calibrator                  &  J0532+0732      &  J0607-0834    &    J0423-0120        \\
Phase Calibrator                     &  J0607-0834      &  J0552+0313    &    J0552+0313        
\enddata
\end{deluxetable*}

%% file: correlator_setup.tex
\begin{deluxetable*}{lccccccc}
\tablecolumns{8}
\tablewidth{0pt}
\tabletypesize{\scriptsize}
\tablecaption{Spectral Window Setup \label{tab:spw}}
\tablehead{
\colhead{} &
\colhead{Central Sky Frequency} & 
\colhead{} &
\colhead{Rest Frequency} &
\colhead{Bandwidth} &
\colhead{Bandwidth} &
\multicolumn{2}{c}{Channel Width} \\
\colhead{Band} &
\colhead{GHz} &
\colhead{Line} &
\colhead{GHz} &
\colhead{MHz} &
\colhead{\kms} &
\colhead{MHz} &
\colhead{\kms}}
\startdata
3\,mm           & 114.968      &    \co{12}{1}{0}       &   115.27120 &   468.75    &  1220    &   0.1221        &  0.32        \\
                & 112.712      &    Continuum           &  \nodata    &   2000.0    &  \nodata &    15.625       &  \nodata     \\
                & 102.420      &    Continuum           &  \nodata    &   2000.0    &  \nodata &    15.625       &  \nodata     \\
                & 100.535      &    Continuum           &  \nodata    &   2000.0    &  \nodata &    15.625       &  \nodata     \\ \hline
1\,mm           & 230.110      &     \co{12}{2}{1}      &  230.53800  &   468.75    &   611    &       0.1221    &  0.16  \\
                & 219.598      &     \co{13}{2}{1}      &  220.39868  &    468.75   &   640    &       0.1221    &  0.17 \\ 
                & 233.370      &      Continuum         &  \nodata    &  2000.0     &  \nodata &       15.625    &  \nodata \\
                & 215.902      &      Continuum         &  \nodata    &  2000.0     &  \nodata &       15.625    &  \nodata \\ \hline
870\,\micron    & 344.262      &     \co{12}{2}{1}      &  345.79599  & 1875.0      & 1634     &       0.488281  &  0.43    \\
                & 330.261      &     \co{13}{2}{1}      &  330.58797  & 1875.0      & 1703     &       0.488281  &  0.44   \\
                & 342.270      &      Continuum         &   \nodata   & 2000.0      & \nodata  &       15.625    &  \nodata \\
                & 332.145      &      Continuum         &   \nodata   & 2000.0      & \nodata  &       15.625    &  \nodata 
\enddata
\end{deluxetable*}

%% file: image_summary_line.tex
\begin{deluxetable*}{cccccccccc}
\tablecolumns{10}
\tablewidth{0pt}
\tabletypesize{\scriptsize}
\tablecaption{Line Image Properties \label{tab:image_summary_line}}
\tablehead{
\colhead{} &
\colhead{Rest Frequency} &
\colhead{FOV} &
\colhead{{\it u-v} range} &
\colhead{} &
\colhead{Beam} &
\colhead{PA} &
\colhead{Velocity Resolution} &
\multicolumn{2}{c}{Noise} \\
\colhead{Transition} &
\colhead{GHz} & 
\colhead{\arcsec} &
\colhead{k$\lambda$} &
\colhead{Weighting} &
\colhead{\arcsec} &
\colhead{\degr} &
\colhead{\kms} &
\colhead{\mJybeam} &
\colhead{K} }
\startdata
\co{12}{3}{2}  & 345.79599    & 18.2 &   16.6 - 496 &  \robzero & $0.54\times0.49$   & 28    & 2~\kms & 2.6  &  0.1    \\ 
\co{12}{2}{1}  & 230.53800    & 27.3 &   16.6 - 496 &  \robzero & $0.54\times0.49$   & 28    & 2~\kms & 3.74 &  0.3    \\ 
\co{12}{1}{0}  & 115.27120    & 54.7 &   16.6 - 496 &  \robzero & $0.54\times0.49$   & 28    & 2~\kms & 5.2  &  1.8     \\ \hline
\co{13}{3}{2}  & 330.58797    & 19.1 &   16.6 - 496 &  \robzero & $0.54\times0.49$   & 27    & 2\kms  & 2.4  &  0.1     
\enddata
\end{deluxetable*}


%% file: image_summary_cont.tex
\begin{deluxetable*}{cccccccc}
\tablecolumns{8}
\tablewidth{0pt}
\tabletypesize{\scriptsize}
\tablecaption{Continuum Image Properties \label{tab:image_summary_cont}}
\tablehead{
\colhead{Wavelength} &
\colhead{Frequency} &
\colhead{} &
\colhead{FOV} &
\colhead{Beam} &
\colhead{PA} &
\multicolumn{2}{c}{Noise} \\
\colhead{mm} &
\colhead{GHz} &
\colhead{Telescope} &
\colhead{\arcmin} & 
\colhead{\arcsec} &
\colhead{\degr} &
\colhead{\uJybeam} &
\colhead{mK}}
\startdata
0.86       & 344.9 & ALMA   &  0.3  & $1.0\times1.0$ & 0 & 54.2  & 0.55 \\
1.3        & 233.3 & ALMA   &  0.45 & $1.0\times1.0$ & 0 & 63.1  & 1.4\\
2.7        & 112.7 & ALMA   &  0.93 & $1.0\times1.0$ & 0 & 41.7  & 4.0\\
13         & 22.46 &  VLA   &  2.0  & $1.0\times1.0$ & 0 & 53.8  & 130.1\\
35         & 8.46  &  VLA   &  5.3  & $1.0\times1.0$ & 0 & 58.3  & 993.8\\
62         & 4.86  &  VLA   &  9.3  & $1.0\times1.0$ & 0 & 16.7  & 862.6
\enddata
\end{deluxetable*}


%% file: iizw40_cont_allbands_nat_uvtaper_1arcsec_v1.tex
\begin{deluxetable}{ccc}
\tablewidth{0pt}
\tablecaption{Radio/Submillimeter Continuum Spectrum of II Zw 40 \label{tab:radio_sed}}
\tablecolumns{3}
\tablehead{
    \colhead{} &
    \colhead{Frequency} &
    \colhead{Flux Density\tablenotemark{a,b}} \\
    \colhead{Wavelength} &
    \colhead{GHz} &
    \colhead{mJy} }
\startdata
   6.2 cm &  4.86 & $  9.4 \pm   1.9$ \\ 
   3.5 cm &  8.46 & $  7.5 \pm   1.5$ \\ 
   1.3 cm & 22.46 & $  7.1 \pm   1.4$ \\ 
   2.7 mm & 112.70 & $  4.7 \pm   0.9$ \\ 
   1.3 mm & 233.38 & $  5.9 \pm   1.2$ \\ 
 870.0 \micron & 344.88 & $  7.7 \pm   1.5$ 
\enddata \tablenotetext{a}{Flux densities derived from matched beam
and uv-coverage images in a 9\arcsec\ diameter aperture. See
Figure~\ref{fig:cont_overview} for the location of the aperture.}
\tablenotetext{b}{The errors are estimated to be 20\% at all
wavelengths except for the 870\micron\ band, based the accuracy of the
flux density scales. The error is estimated to be 10\% at 870\micron\
because we were able to cross-check the flux calibrator brightness. See
\S~\ref{sec:data} for more details. }  \end{deluxetable}

%% file: iizw40_gmc_props_final_v1_astroph.tex
\clearpage
\begin{turnpage}
\begin{deluxetable*}{crrrrrrrrrrrrrllcc}
\tablewidth{0pt}
\tabletypesize{\scriptsize}
\tablecaption{Properties of Giant Molecular Clouds Within \iizw \label{tab:gmc_props}}
\tablecolumns{18}
\tablehead{
    \colhead{} & 
    \colhead{RA} & 
    \colhead{Dec} & 
    \colhead{V} & 
    \multicolumn{3}{c}{$L_{CO(3-2)}$} & 
    \multicolumn{3}{c}{$L_{CO(2-1)}$} & 
    \multicolumn{3}{c}{$L_{CO(1-0)}$} & 
    \colhead{$\sigma_v$} & 
    \colhead{$R$} & 
    \colhead{$M_{v}$} & 
    \colhead{$T_{b}(3-2)$} & 
    \colhead{$T_{b}(2-1)$}  \\ 
    \colhead{Label} &
    \colhead{J2000} &
    \colhead{J2000} &   
    \colhead{\kms} &
    \multicolumn{3}{c}{\Kkmspc} &
    \multicolumn{3}{c}{\Kkmspc} &
    \multicolumn{3}{c}{\Kkmspc} &
    \colhead{\kms} &
    \colhead{pc} &
    \colhead{$10^5 M_\odot$} &
    \colhead{K}&
    \colhead{K} \\
    \colhead{(1) } &
    \colhead{(2)} &
    \colhead{(3)} &   
    \colhead{(4)} &
    \multicolumn{3}{c}{(5)} &
    \multicolumn{3}{c}{(6)} &
    \multicolumn{3}{c}{(7)} &
    \colhead{(8)} &
    \colhead{(9)} &
    \colhead{(10)} &
    \colhead{(11)} &
    \colhead{(12)}}
\startdata
A &  5:55:42.268 &  3:23:27.45 & 769.0 & 3600 & $\pm$ & 400 & 1100 & $\pm$ & 200  & 6600 & $\pm$ & 3400 & 3.08 $\pm$ 0.02 &  $<$11.4 & $<$1.1 &  0.9 & 1.0   \\ 
B &  5:55:42.572 &  3:23:31.57 & 738.5 & 3000 & $\pm$ & 300 & 1900 & $\pm$ & 400  & 5600 & $\pm$ & 2900 & 4.19 $\pm$ 0.04 &  $<$12.8 & $<$2.3 &  0.5 & 0.8   \\ 
C &  5:55:42.620 &  3:23:32.13 & 754.7 & 18700 & $\pm$ & 2000 & 18300 & $\pm$ & 3700  & 34500 & $\pm$ & 17600 & 5.58 $\pm$ 0.07 &  $<$7.2 & $<$2.3 &  1.3 & 2.5   \\ 
D &  5:55:42.654 &  3:23:32.30 & 746.1 & 22900 & $\pm$ & 2400 & 39500 & $\pm$ & 7900  & 42200 & $\pm$ & 21500 & 4.92 $\pm$ 0.05 &  34.6 $\pm$ 1.2 &  4.6 $\pm$ 0.2 & 1.4 & 2.8   \\ 
E &  5:55:42.733 &  3:23:30.18 & 742.1 & 4100 & $\pm$ & 400 & 2900 & $\pm$ & 600  & 7500 & $\pm$ & 3800 & 2.28 $\pm$ 0.02 &  15.9 $\pm$ 3.8 &  0.4 $\pm$ 0.1 & 0.7 & 1.5   \\ 
F &  5:55:42.790 &  3:23:33.56 & 730.1 & 6100 & $\pm$ & 600 & 4600 & $\pm$ & 900  & 11300 & $\pm$ & 5800 & 2.99 $\pm$ 0.03 &  38.8 $\pm$ 0.7 &  1.9 $\pm$ 0.1 & 0.7 & 1.1   \\ 
G &  5:55:42.808 &  3:23:32.61 & 763.8 & 20700 & $\pm$ & 2100 & 26900 & $\pm$ & 5400  & 60900 & $\pm$ & 31100 & 8.79 $\pm$ 0.04 &  32.5 $\pm$ 0.5 &  13.7 $\pm$ 0.3 & 0.9 & 2.2   \\ 
H &  5:55:42.891 &  3:23:32.03 & 761.4 & 4300 & $\pm$ & 400 & 3000 & $\pm$ & 600  & 7900 & $\pm$ & 4000 & 3.13 $\pm$ 0.03 &  24.4 $\pm$ 0.9 &  1.3 $\pm$ 0.1 & 0.7 & 1.2   \\ 

\enddata 
\tablecomments{Column (1): Cloud label. Columns (2) and (3): Right
  Ascension and Declination of the luminosity weighted center of the
  cloud in J2000. Column (4): Mean luminosity-weighted velocity of
  cloud in the LSRK frame. Column (5): Measured \cothree\
  luminosity. Column (6): Measured \cotwo\ luminosity. Column (7):
  Estimate of \coone\ luminosity based on observed \cothree\
  luminosity and \rthreeone\ ratio (see \S\ref{sec:meas-cloud-prop}
  for details). Column (8): Linewidth of cloud derived using \cothree\ data. Column (9): Size of cloud derived using \cothree\ data. (see \S\ref{sec:meas-cloud-prop} for details). Column (10): Virial mass of cloud derived using \cothree\ data (see \S\ref{sec:meas-cloud-prop} for details). Column (11) and Column (12): Peak \cothree\ and \cotwo\ temperature of the cloud.}
\end{deluxetable*}
\end{turnpage}
\clearpage
\global\pdfpageattr\expandafter{\the\pdfpageattr/Rotate 90}